\documentclass[final,leqno]{siamltex}
\usepackage{cmmib57}

\usepackage{amsmath,amssymb}
\usepackage{exscale}

\usepackage{graphicx}

\newtheorem{Theorem}{Theorem}
\title{Periodically generated propagating pulses\thanks{This research was supported by 
the Spanish MCyT grant BFM2002-04127-C02-01, and by the European Union under
grant HPRN-CT-2002-00282.  Received by the editors of SIAM J. Appl. Math.
on  . Manuscript number 043494-1.} }

\author{L. L. Bonilla\thanks{Escuela Polit{\'e}cnica
Superior, Universidad Carlos III de Madrid, Avenida de la Universidad 30, 
28911 Legan{\'e}s, Spain (bonilla@ing.uc3m.es).}
\and M. Kindelan\thanks{Escuela Polit{\'e}cnica
Superior, Universidad Carlos III de Madrid, Avenida de la Universidad 30, 
28911 Legan{\'e}s, Spain (kinde@ing.uc3m.es).} \and
J. B. Keller\thanks{Departments of Mathematics and Mechanical Engineering,
Stanford University,
Stanford, CA 94305, USA (keller@math.stanford.edu)} } 
\begin{document}

\date{\today}
\maketitle
\vspace{-1.2in}
%\slugger{siap}{2003}{}{}{1--27}
\vspace{.9in}

\setcounter{page}{1}
\renewcommand{\thefootnote}{\arabic{footnote}}
\renewcommand{\theequation}{\arabic{section}.\arabic{equation}}
\newcommand{\fin}{\newline \rule{2mm}{2mm}}
\def\RR{\mathbb{R}}
\def\pRR{\mathbb{R}}
\def\ZZ{\mathbb{Z}}

\begin{abstract}
Certain equations with integral constraints have as solutions time-periodic
pulses of a field-like unknown while a current-like
unknown oscillates periodically with time. A general asymptotic theory of
this phenomenon, the generalized Gunn effect, has been found recently.
Here we extend this theory to the case of nonlinearities having only one
stable zero, which is the case for the usual Gunn effect in n-GaAs. 
Our ideas are presented in the context of a simple scalar model
where the waves can be constructed analytically and explicit expressions for
asymptotic approximations can be found.
\end{abstract}
%\pacs{5.45.-a}
%\maketitle
%\begin{multicols}{2}
%\narrowtext

%\begin{AMS}
34E15, 92C30. \hspace{2cm} Date: \today
%\end{AMS}

\pagestyle{myheadings}
\thispagestyle{plain}
\markboth{L.~L.~BONILLA, M.~KINDELAN AND J.~B.~KELLER}{PERIODICALLY
GENERATED PROPAGATING PULSES}

\setcounter{equation}{0}
\section{Introduction}\label{sec:intro}
The Gunn effect is the periodic oscillation of the current in a passive external circuit
attached to a dc-voltage biased semiconductor whose electron drift velocity has a single
maximum as a function of the electric field (and therefore the curve of electron velocity
versus electric field has negative slope for field values on a certain interval, a fact called 
{\em negative differential mobility}) \cite{shaw79,vol69}. During each period 
of the oscillation, a pulse of the electric field is created at the injecting
contact, moves through the semiconductor, and is annihilated at the receiving
contact. While originally observed in bulk n-GaAs samples, similar current oscillations, 
mediated by pulse dynamics in dc voltage biased semiconductors, have been found in many 
materials, several of which lack negative differential mobility \cite{bonc74}. Instead, other 
processes (impact ionization at impurities \cite{teit89}, nonlinear capture
coefficients \cite{samu95}, nonlinear recombination processes, etc) may be
responsible for a current vs.\ local electric field characteristic curve
displaying a local maximum followed by a region of negative slope (negative
differential conductivity). 

Propagation of pulses occurs in many systems of interest in Biology, Physics, 
\ldots: morphogen pulses or spikes in activator-inhibitor reaction-diffusion systems 
modeling cell development or chemical reactors \cite{morpho3,doe00,hem98,STWW}, 
propagation of nerve impulses along myelinated or unmyelinated fibers 
\cite{kee98,RK73,kee87}, pulse propagation through cardiac cells \cite{kee98}, 
calcium release waves in living cells \cite{bug97}, semiconductor superlattices 
\cite{bon02,wac02} and oscillatory instabilities of the current in bulk
semiconductors with an N-shaped current--field characteristic \cite{bonc74,shaw79,vol69}.
These distributed systems can be spatially discrete or continuous and can be
described by a variety of model equations. Sometimes a pulse is created from
an appropriate initial condition and it reaches a stable shape, moving
uniformly until it arrives at a boundary. Sometimes understanding pulse
dynamics is the key to describing a more complicated evolution of the system. A
good example is the Gunn effect in bulk semiconductors. 

\begin{figure}
\begin{center}
\includegraphics[width=8cm,angle=270]{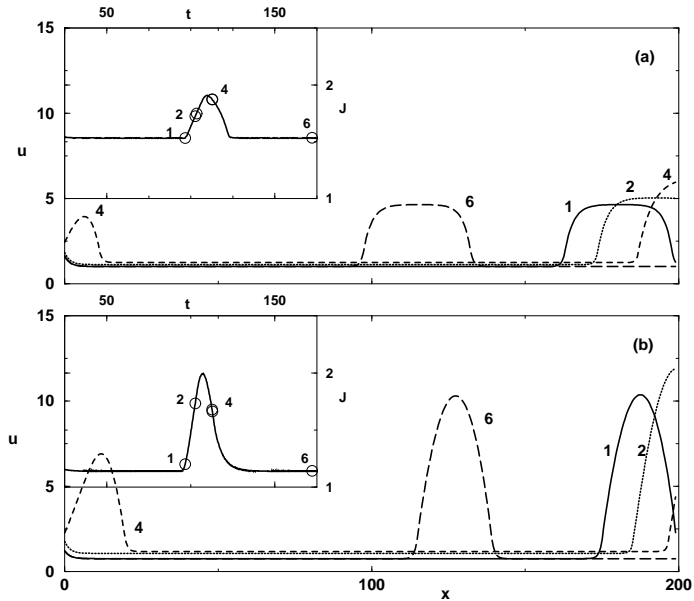}
\caption{Self-sustained oscillations of the `current' $J(t)$ (see inset, which shows one 
period of $J(t)$) mediated by pulses 
of the `field' $u(x,t)$, which are solutions of the model described in Section 2. The pulse 
profiles correspond to the times marked in the inset. Case (a): bistable source. Case (b): 
source with a single stable zero. }
\label{fig9}
\end{center}
\end{figure}

While Gunn-like instabilities have been known for a long time, only
recently have pulse annihilation and creation at boundaries
been studied by asymptotic methods \cite{HB92,BCGR97,BC96}. These
theories treated the case in which the relevant nonlinear source term
has two stable zeros. Then there are stable wavefronts joining these two
zeros, and a moving pulse is a flat region of high field bounded by two
wavefronts. The pulse changes its size if its leading and trailing wavefronts
move at different speeds. Fig.~\ref{fig9} shows the time-periodic oscillation of
a `current' $J(t)$ accompanied by the repeated generation and motion of flat-top pulses of an 
`electric field' $u(x,t)$, which are solutions of model equations described in Section 
\ref{sec:ssm}. Fig.~\ref{fig9}(a) corresponds to the case of a bistable nonlinear source. 
The asymptotic analysis of the Gunn effect is based on the dynamics of such pulses 
\cite{BCGR97}. However, the source term (velocity-field characteristic curve) in very 
relevant materials, such as bulk n-GaAs \cite{shaw79}, wide-miniband GaAs/AlAs 
superlattices \cite{butt77,hofb96}, and semi-insulating GaAs \cite{samu95}, does not 
have two stable zeros. Instead, the nonlinearity has a single stable zero, so the previous 
theories, based on two moving wavefronts, are not valid. Fig.~\ref{fig9}(b) shows the self-sustained
oscillations corresponding to this case. Can we find an asymptotic theory of pulse 
mediated oscillations in this case? The answer is yes, as we show in this paper.

To emphasize that our analysis applies to a class of models, we shall analyze
a simpler problem than that of the Gunn effect. We will study a nonlinear
reaction-diffusion-convection equation with an integral constraint
\cite{BC96} and a piecewise linear source term. Then the pulses can be
constructed analytically, their size can change as they move, and explicit 
expressions for the asymptotic approximations can be found. 
Such a construction was used by Rinzel and
Keller for the FitzHugh---Nagumo equation \cite{RK73}. For Kroemer's model, a
piecewise linear electron velocity was used to calculate the exact form and speed of
a steadily propagating pulse \cite{butc66}. 

The outline of the paper is as follows. Section \ref{sec:ssm}
presents our simplified model. It also reviews the key ideas of
previous asymptotic theories, valid for an N-shaped nonlinearity with
two stable zeros. Section \ref{sec:univ} discusses the construction of stationary 
solutions, and the kinematics of wave fronts, in the limit of long samples. When the 
nonlinearity is piecewise linear, the pulses can be
found analytically. Section \ref{sec:1pulse} discusses the dynamics of a
single pulse moving from the injecting to the receiving boundary. We show that
the pulse changes its form and speed adiabatically, following the instantaneous
value of the current. In Section \ref{sec:dynamics}, we complete our description of
the oscillations by explaining what happens when the pulse reaches the
receiving boundary, and how a new pulse is created at the injecting boundary.
Section \ref{sec:conclusions} contains our conclusions. The Appendices are
devoted to technical matters.

\setcounter{equation}{0}
\section{Simple scalar model}
\label{sec:ssm}

The model consists of a one-dimensional nonlinear parabolic
equation for $u(x,t)$ (the ``electric field'') with an unknown forcing term
$J(t)$ (the ``current''). There are also an integral constraint (the ``voltage
bias condition''), as well as boundary and initial conditions:
\begin{eqnarray}
{\partial u\over\partial t} + K\, {\partial u\over\partial x}
 = {\partial^{2} u\over\partial x^{2}} + J -  g(u)\,, \quad 0<x<L, \label{eq1}\\
{1\over L}\, \int_0^L u(x,t)\, dx = \phi, \label{bias}\\
u(0,t) =
%u(L,t) =
\rho\, J(t) , \quad\quad \rho > 0,
\qquad {\partial u\over\partial x}(L,t) = 0 \label{bc}\\
u(x,0) = f(x) \geq 0, \quad\quad 0<x<L.\label{ic}
\end{eqnarray}
Here $K$, $\rho$ and $\phi$ are positive constants. 

This model becomes the Kroemer model of the Gunn effect if the advection constant $K$ 
in Eq.\ (\ref{eq1}) is replaced by $g(u)$ \cite{bon91,BHV94,onset,HB92,shaw79,vol69}. 
For the Kroemer model, existence and 
uniqueness of solutions have been studied by Liang \cite{Liang}. Furthermore,
linear stability of the stationary and moving pulse solutions have been analyzed in 
\cite{BHV94}; see also \cite{shaw79,vol69}. A bifurcation analysis of the 
self-sustained oscillations due to pulse dynamics near critical values of $\phi$ can
be found in \cite{onset}. Asymptotic analyses of the Gunn effect for the Kroemer 
model with a bistable $g(u)$ can be found in \cite{HB92,BCGR97}. 

\subsection{Bistable source  $g(u)$} \label{sec:nosat}
The simplest case to consider is that of an N-shaped nonlinearity $g(u)$ (the
`velocity--field characteristic') for $u\geq 0$, with a local maximum $g_M =
g(u_M)$, $u_M >0$, followed by a local minimum $g_m = g(u_m)>0$, $u_m >u_M$.
Then $J-g(u)$ may have up to three positive zeroes for $J>0$,
namely $u_1(J)<u_2(J)<u_3(J)$. For large enough $L$ and $K$,
$g_M/u_M<\rho <g_m/u_m$, and for $\phi$ in a certain subinterval of
$(u_M,u_3(g_M))$, there  are stable time periodic solutions of
(\ref{eq1}) - (\ref{bc}) of Gunn type. As shown in Fig.~\ref{fig9}(a), while 
$J(t)$ oscillates periodically, pulses of $u(x,t)$ are created at $x=0$, move
towards $x=L$, and disappear there. 

The analysis of the model is simple in the asymptotic limit
\begin{eqnarray}
0<\epsilon \equiv {1\over L} \ll 1. \label{limit}
\end{eqnarray}
In this limit (\ref{eq1}) - (\ref{bias}) may be
written as
\begin{eqnarray}
{\partial u\over\partial s} + K\, {\partial u\over\partial y} =
\epsilon {\partial^{2} u\over\partial y^{2}} + {J -  g(u)\over
\epsilon}\,,\label{eq2}\\
\int_0^1 u(y,s)\, dy = \phi, \label{scaled.bias}
\end{eqnarray}
where
\begin{eqnarray}
y = \epsilon x,\quad s = \epsilon t. \label{slow.scales}
\end{eqnarray}

Eq.\ (\ref{eq2}) is a parabolic equation with fast reaction and slow
diffusion terms. As $\epsilon\to 0+$, $u(y,s)$ is typically a piecewise
constant function taking on the {\em order one values} $u_1(J)$ or $u_3(J)$
in intervals of length $y=O(1)$. The extrema of these intervals are typically
moving internal layers. These layers are important because they bound pulses,
and the self-sustained oscillation we want to describe is due to recycling
and motion of pulses at the boundaries. A pulse is a region where $u$ is
$u_3(J)$, separated by moving wavefronts from two other regions where $u$ is
$u_1(J)$. There are two wavefronts bounding the pulse. In the backfront
$u$ increases from $u_1(J)$ to $u_3(J)$; this front moves with a
speed $c_+(J)$. The forefront moves at speed $c_-(J)$, and in it 
$u$ decreases from $u_3(J)$ to $u_1(J)$. Forefront and backfront
are heteroclinic trajectories connecting the two saddles $(u_1(J),0)$ and
$(u_3(J),0)$ in an appropriate phase plane $(u,du/d\chi)$, where $\chi =
[y-Y(s)]/\epsilon$ is a moving coordinate ($\chi = 0$ at the wavefront and
$dY/ds = c_{\pm}(J)$). The instantaneous value of $J(s)$ is determined by
using the integral condition (\ref{scaled.bias}). Typically $J$ obeys the
simple equation
\begin{equation}
{dJ\over ds} = A(J)\, [n_+ c_+(J) - n_- c_-(J)], \label{eq_J}
\end{equation}
where $A(J)>0$ is a known function of $J$, and $n_+$ and $n_-$ are the
numbers of wavefronts with increasing and decreasing $u$ profiles,
respectively. For high-field domains, $n_+ - n_- = 0, 1$
\cite{BC96,BCGR97}. The fixed points of Eq.\ (\ref{eq_J})
correspond to the equal area rule
\begin{equation}
\int_{u_{1}}^{u_{3}} [g(u) - c]\, du = 0
\end{equation}
if $n_+ = n_-$, or to possible plateaus in the shape of $J(s)$ otherwise.
Many questions on the stability of the pulses and their evolution can be
simply answered by analyzing Eq.\ (\ref{eq_J}) and using the asymptotic procedure
explained in \cite{BC96,BCGR97}. The description of pulse creation and
annihilation at the boundaries requires a finer analysis, as explained in
\cite{BC96,BCGR97}.

\subsection{Saturating source} \label{sec:sat}
If $g(u)$ saturates, i.e.\  $g(u)\to$constant as $u\to\infty$, no $u_3(J)$ exists and the 
previous construction is no longer possible. What is the correct asymptotic
description of a pulse in this case? Let us anticipate the answer here. As Fig.~\ref{fig9}(b)
shows, a pulse is a traveling wave whose profile has a single maximum, and tends
to $u= u_1(J)$ as $x\to\pm\infty$. The pulse is bounded by a leading
front (forefront, moving at speed $c_-$) and a trailing front (backfront,
moving at speed $c_+$). Two parameters uniquely determine the pulse: its
maximum height, $u_m$, and the instantaneous value of $J$. Given $u_m$ and
$J$, we can find the wavefronts enclosing the pulse by simple phase plane
constructions. Consider the phase plane $(u,du/d\chi)$,
$\chi=x-X(t)$, where $X(t)$ is the instantaneous position of a wavefront and
$dX/dt$ its speed. There is a unique speed $c_+= c_+(J,u_m)$ for which a
separatrix issuing from the saddle $(u_1(J),0)$ on the upper half plane
reaches the $u$ axis at $(u_m,0)$. This separatrix constitutes the
backfront of the pulse, and a similar construction supplies its forefront
moving at speed $c_-(J,u_m)$. In general, $c_+ \neq c_-$, which implies that
our pulse changes its size as it moves. How do we characterize the dynamics
of pulses?

Suppose that there is a single pulse moving in the sample. We need
to determine the instantaneous values of $J$ and $u_m$, for they
completely characterize the pulse. The pulse width, $l$, changes
as $dl/dt = c_- - c_+$. On the other hand, $l$ may be determined
by a line integral on the corresponding phase plane trajectories
which form the pulse. Then $l=l(J,u_m)$. The dc bias condition
yields a connection between $u_m$ and $J$, $u_m = U(J)$. Then
the pulse width is a function of $J$ only, $\varphi(J) = l(J,U(J))$.
Therefore since $dl/ds = \varphi'(J)\, dJ/ds = c_{-}- c_{+}$, we get
\begin{equation}
{dJ\over ds} = {c_{+}(J,U(J)) - c_{-}(J,U(J))\over -\varphi'(J)},
\label{eq_J1}
\end{equation}
Typically the fixed point of this equation, $J=J^*$, is such that
$c_+ =c_-$ is a globally stable solution, so that $J$ tends
exponentially fast to $J^*$. The corresponding pulse moves
steadily without changing its size. This pulse is the homoclinic
orbit in the phase plane, usually given by an equal-area rule and
exhaustively studied by previous authors. Notice that the present
construction explains why this steadily moving pulse is stable,
thereby clarifying an old issue at the heart of the Gunn effect \cite{vol69,bon91}.
When the pulse reaches the receiving contact, a different stage
of the Gunn oscillation begins. This stage and others needed to
fully describe the Gunn oscillation will be explained later.

A subtle point is the following. Due to the integral condition (\ref{bias}),
the pulse height and width [in the $(y,s)$ scales] are $O(\epsilon^{
-{1\over 2}})\gg 1$ and $O(\epsilon^{{1\over 2}})$ respectively,
while outside the pulse, $u=u_1 = O(1)$ and $J=O(1)$. Thus our
leading order approximation for $u(y,s)$ is not uniformly of the
same order in space. Successive approximations of a single pulse
%(which we will need in our analysis)
lead to the following ansatz
for $u$:
$$ u(y,s;\epsilon) \sim u^{(0)}(y,s;\epsilon) + \epsilon\,
u^{(1)}(y,s;\epsilon).$$
Here each $u^{(j)}(y,s;\epsilon)$ may be of different order in $\epsilon$
for different values of the space and time variables. However we shall
impose that
$${u^{(1)}\over u^{(0)}} = O(1),$$
uniformly in $y$ and $s$ as $\epsilon\to 0+$. This situation results in a
changing (self-adjusting) time scale for the evolution of $J$ described
by (\ref{eq_J1}). See Ref. \cite{BCS00} for the description of
a similar situation in combustion theory.

\setcounter{equation}{0}
\section{Boundary layers and wavefronts}
\label{sec:univ}
\begin{figure}
\begin{center}
\includegraphics[width=8cm]{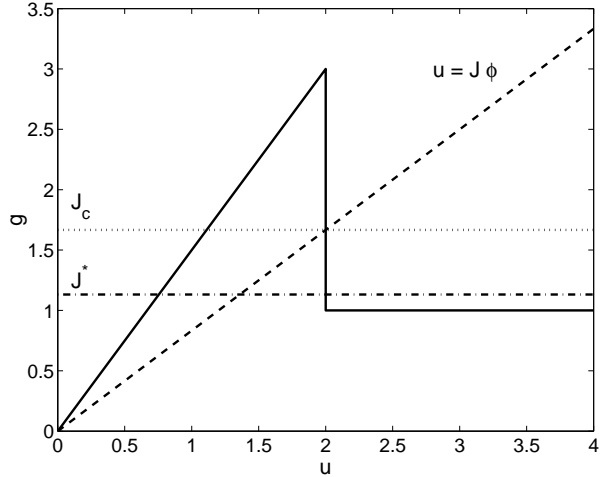}
\caption{The nonlinearity function $g(u)$ (solid), and the curve
corresponding to the boundary conditions: $J=u/\rho$ (dashed).
The parameter values are $\phi = 1.6$, $L=200$, $K=2.0$, $u_M =
2.0$, $\beta = 1.5$, $\alpha = 1.0$, $\rho = 1.2$. }
\label{fig1}
\end{center}
\end{figure}

The model (\ref{eq1}) - (\ref{ic}) was introduced in order to argue
that the asymptotics of the Gunn effect is universal within a class
of models~\cite{BC96}. The nonlinearity $g(u)$ was originally N-shaped:
it had three branches for $u>0$ (with a maximum at $u_M>0$ and a minimum at
$u_{\text{min}}>u_M$ such that $g(u_M)>g(u_{\text{min}})>0$, with $g(u)
\to\infty$ as $u\to\infty$). In the present paper, we shall assume that $g(u)$ is a
smooth function having a single maximum at $u_M >0$, $g(u_M)=g_M > 0$,
such that $g'(0) = \beta > 0$, and lim$_{u\to\infty} g(u) = \alpha
\in (0,g_M)$. To obtain explicit analytic expressions, we shall use
a piecewise linear version of $g(u)$,
\begin{eqnarray}
g(u) = \beta u\, \theta(u_M-u) + \alpha\,\theta(u-u_M)\,,
\label{g}
\end{eqnarray}
where $\theta(x)$ is the Heaviside unit step function. See figure
\ref{fig1} where we have also shown a typical
straight line $J=u/\rho$ which gives the value of $u$ at the
%boundaries.
left boundary. Notice that the straight line intersects $g(u)$ at
$(u_M,\rho u_M)$, with $\alpha <\rho u_M < g_M$. We want to find
stable solutions $(u(x,t),J(t))$ of Eqs.~(\ref{eq1}) and
(\ref{bias}) under the boundary conditions (\ref{bc}) and the
initial condition (\ref{ic}) for appropriate positive values of
the bias $\phi$, in the asymptotic limit $L\to\infty$.

\subsection{Outer limit and boundary layers}
If $0<\phi<u_M$, there is a single stationary solution of Eqs.~(\ref{eq1})
- (\ref{bc}) which is easily constructed. Let $y = x/L \equiv \epsilon x$,
$0<\epsilon\ll 1$.
Introducing this scaling in Eq.~(\ref{eq1}) yields the leading order equation
\begin{eqnarray}
J - g(u) = 0\,,\label{eq3}
\end{eqnarray}
outside the boundary layers at $y = 0,1$. For $0<J<g_M$ and $0<y<1$,
(\ref{eq3}) has the solutions $u_1(J)<u_2(J)$. Of these we may choose
$u= u_1(J)$, $0<y<1$, as the outer limit. Inserting it into (\ref{bias}),
we find $u_1(J) + O(\epsilon) = \phi$ and then (\ref{eq3}) yields
\begin{eqnarray}
 J = g(\phi) \,.   \label{eq4}
\end{eqnarray}
For the piecewise linear $g(u)$, $J= \beta \, \phi$, and $u_1(J) \, = \, J \, / \beta$.

The boundary layers at $y=0,1$ are solutions of the problems
\begin{eqnarray}
{\partial^{2} U\over\partial \xi^{2}} \mp K {\partial U\over\partial \xi}
+ J - g(U) = 0\,,\quad\quad 0 < \xi < \infty, \label{eq5}\\
U(0) = \rho J, \quad U(\infty) = u_1(J) .\label{eq6}
\end{eqnarray}
Here $\xi = x = y/\epsilon$ for the injecting boundary layer at
$y=0$, and $\xi = L-x = (1-y)/\epsilon$ for the receiving boundary
layer at $y=1$. The minus (resp.\ plus) sign in (\ref{eq5}) corresponds
to the injecting (resp.\ receiving) boundary. Clearly the shape of the
unique solution of (\ref{eq5}) and (\ref{eq6}) (for $\xi = x$) depends
on whether $J$ is smaller or larger than $J=J_c = u_M/\rho$. When
$0<J<J_c$, the boundary layer profile monotonically decreases from $u =
\rho J$ to $u_1 = J/\beta$. For the piecewise linear $g(u)$, we have
\begin{eqnarray}
U(\xi) = {J\over\beta} + \left(\rho J - {J\over\beta}\right)
\exp\left(-{\sqrt{K^{2}+4\beta}\mp K\over 2}\,\xi\right) ,\label{eq7}\\
\int_0^{\infty} (U - u_1)\, d\xi = {2\,\left(\rho J- {J\over\beta}\right)
\over \sqrt{K^{2}+4\beta}\mp K}\,.\label{eq8}
\end{eqnarray}
However, if $J>J_c$, the boundary layer profile reaches a maximum
before decreasing to $u_1 = J/\beta$.
Numerical simulations show that in this case, the stationary
solution of the model becomes unstable to Gunn-type oscillations.
Given (\ref{eq4}), this occurs for $\phi >
u_1(J_c)$.

\subsection{Wavefronts}

\subsubsection{Bistable source}
Let us first review how to compute the wavefronts when $g(u)$ is N-shaped.
Then (\ref{eq3}) has three solutions $u_1(J)<u_2(J)<u_3(J)$ for $g(u_m)<J
<g(u_M)$. As explained in Ref.~\cite{BC96}, the building blocks of the
Gunn-oscillation asymptotics are wavefronts connecting $u_1(J)$ and $u_3(J)$.
These wavefronts adjust themselves instantaneously to the value of $J$, as
this unknown evolves on a slower time scale (see below). A wavefront centered
at $x=X_{\pm}(t)$ is a monotone function of $\chi = x-X_{\pm}(t)$ such that
\begin{eqnarray}
 u(-\infty;c_+) = u_1(J),\quad u(+\infty;c_+) = u_3(J)
\quad\mbox{and}\nonumber\\
 u(-\infty;c_-) = u_3(J),\quad
u(+\infty;c_-) = u_1(J) .\nonumber
\end{eqnarray}
For the simple model used here, there is a relation between the wavefronts
$u(\chi;c_+)$ and $u(\chi;c_-)$:

{\Theorem Let $u(\chi;c_{\pm})$ be the wavefront satisfying
\begin{eqnarray}
{d^{2} u\over d\chi^{2}} - (K-c_{\pm})\, {du\over d\chi} + J -  g(u)
= 0\,,\label{fronts}\\
u(-\infty;c_{+}) = u_{1}(J),\quad\quad u(+\infty;c_{+}) =
u_{3}(J),\label{bc+}\\
u(-\infty;c_{-}) = u_{3}(J), \quad\quad u(+\infty;c_{-}) = u_{1}(J),\label{bc-}
\end{eqnarray}
where $\chi = x - X_{\pm}(t)$ ($X_{\pm}$ is the position of the front at
time $t$, determined by imposing $u(0)=u^0$. $dX_{\pm}/dt = c_{\pm}$). Then
we have
\begin{eqnarray}
u(\chi;c_{-}) = u(-\chi;c_{+}), \quad\quad c_{+} + c_{-} = 2K.
\label{th1}
\end{eqnarray}
}

The proof is evident. This theorem shows that we only need to construct
$u(\chi;c_+)$ and find $c_+$ in order to have $u(\chi;c_-)$ and $c_-$.

\subsubsection{Saturating source}
Let now $g(u)$ be a function with only two branches such as (\ref{g}). A
wavefront is the only monotone trajectory connecting $(u_1(J),0)$ and a given
point on the $u$ axis $(u_m,0)$. There is again a symmetry result for these
wavefronts:

{\Theorem Let $u(\chi;c_{+})$ be the wavefront satisfying (\ref{fronts}),
$\partial u/\partial \chi >0$ for $-\infty<\chi \equiv x-X_+(t)<\chi_m$,
and
\begin{eqnarray}
u(-\infty;c_{+}) = u_{1}(J),\quad\quad u(0;c_+) = u^0,\nonumber\\
u(\chi_{m};c_{+}) = u_m,\quad\quad
{\partial u\over\partial\chi}(\chi_{m};c_{+}) = 0 .
\label{bc++}
\end{eqnarray}
Here $dX_+/dt = c_+$, and $2\chi_m = l(J,u_m)>0$ is a function of $J$
and $u_m$. Then the wavefront satisfying (\ref{fronts}),
$\partial u/\partial \chi < 0$ for $-\chi_m<\chiÝ\equiv x - X_-(t)
<\infty$, $c_- = dX_-/dt$, and
\begin{eqnarray}
u(-\chi_{m};c_{-}) = u_m,\quad\quad
{\partial u\over\partial\chi}(-\chi_{m};c_{-}) = 0,\nonumber\\
 u(0;c_-) = u^0,\quad\quad u(+\infty;c_{-}) = u_{1}(J),
\label{bc--}
\end{eqnarray}
is such that (\ref{th1}) holds.
}

Again the proof is immediate. 

\begin{figure}
\begin{center}
\includegraphics[width=8cm]{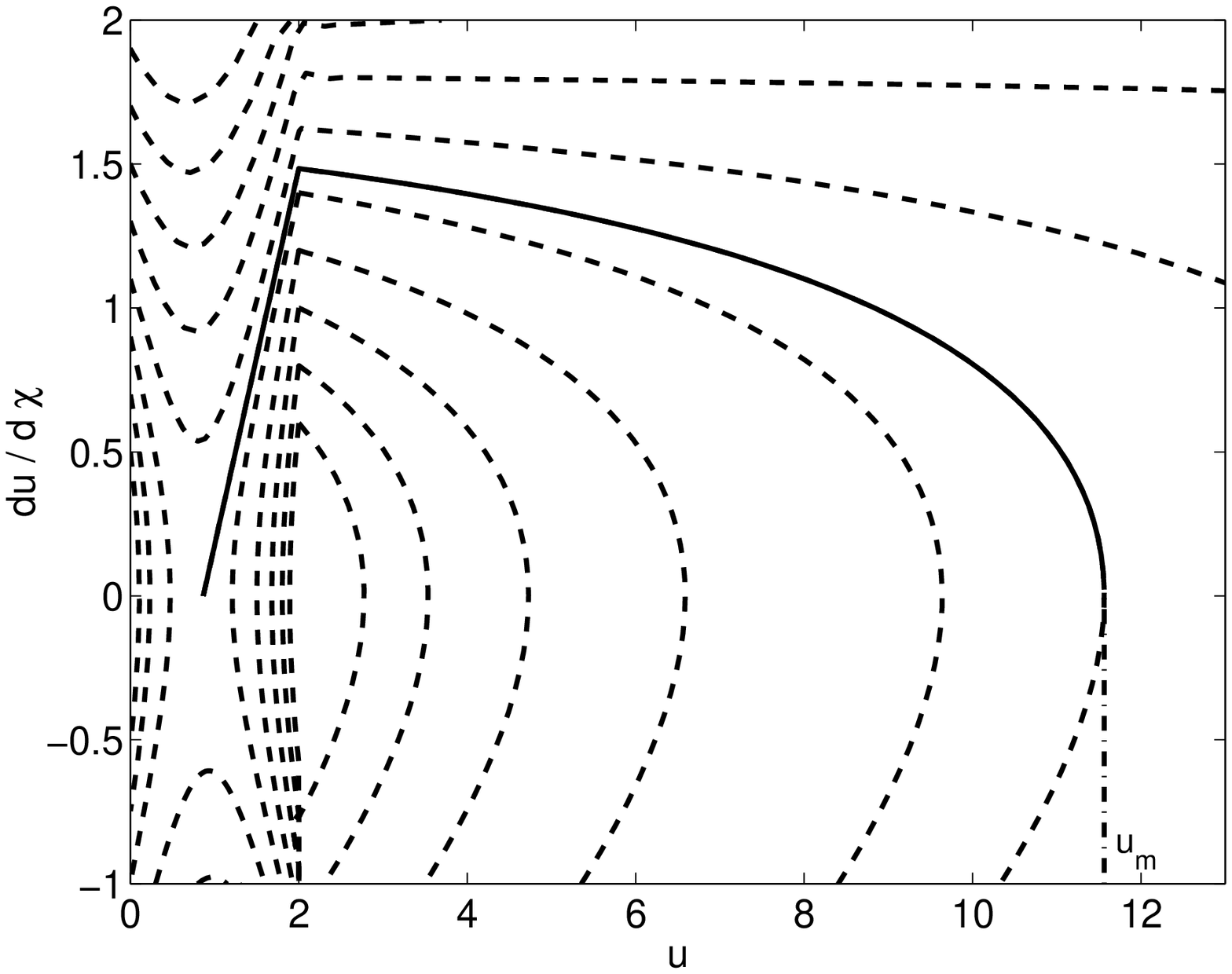}
\caption{ Phase plane $(u,du/d\chi)$ and trajectory corresponding
to $u(\chi;c_+)$ for $J=1.3$, $c_+ = 1.8359$. Other parameter values
as in Figure \ref{fig1}. }
\label{fig5}
\end{center}
\end{figure}

\begin{figure}
\begin{center}
\includegraphics[width=8cm]{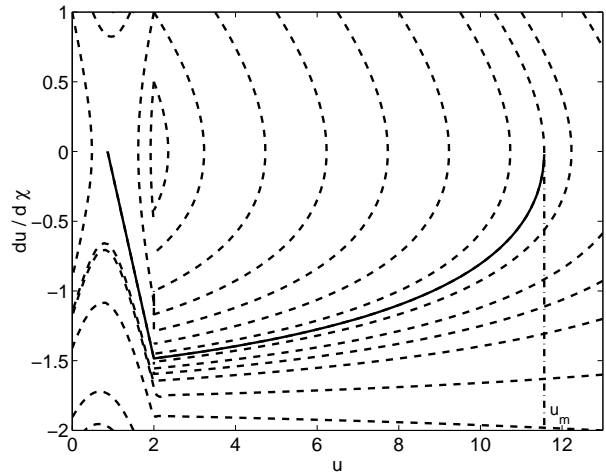}
\caption{ Phase plane $(u,du/d\chi)$ and trajectory corresponding
to $u(\chi;c_-)$ for $J=1.3$, $c_- = 2.1641$. Other parameter values
as in Figure \ref{fig1}. }
\label{fig6}
\end{center}
\end{figure}

Let us now choose a certain $u_m = U(J)$ for each $\phi >u_M$ so that the bias
condition (\ref{bias}) holds for a pulse made out of:
\begin{itemize}
\item a backfront $u(\chi;c_+)$ at $x=X_+(t)$, $\chi =x-X_+(t)$; and
\item a forefront $u(-\chi;c_+)$ at $X_- = X_+ + 2\chi_m$. Now
$\chi =x-X_-(t)$.
\end{itemize}
The pulse $u(\chi;c_+)$ can be constructed
explicitly for the piecewise linear $g(u)$ of
(\ref{g}). If we choose $u^0 = u_M$, the
solution of Eqs.~(\ref{fronts}) and (\ref{bc+})
which is continuous and has a continuous first
derivative at $\chi = 0$ is
\begin{eqnarray}
&& u(\chi;c_+) = u_1(J) + [u_M - u_1(J)]\,
e^{\lambda_{+} \chi}\quad\chi<0\,,\label{<0}\\
&& u(\chi;c_+) = u_M + {J-\alpha\over K-c_{+}}\,\chi
+ B_+\,\left[ e^{(K-c_{+}) \chi} - 1\right]\quad
\chi>0\,.\label{>0}
\end{eqnarray}
Here $u_1(J) = J/\beta$, and
\begin{eqnarray}
\lambda_{+} &=& {K - c_{+}\over 2} + \sqrt{\left(
{K - c_{+}\over  2}\right)^{2} + \beta }\,,
\label{lambda}\\
B_{+} &=& {1\over K - c_{+}}\,\left[ \lambda_{+}
[u_M - u_1(J)] - {J-\alpha\over K -
c_{+}}\right]\,,  \label{B+}\\
\chi_m &=& {1\over K - c_{+}}\,\ln\left[ - {J-
\alpha\over (K -  c_{+})^{2}\, B_{+}}
\right]\nonumber\\
&=& - {1\over K - c_{+}}\,\ln\left[1 -
{\lambda_{+} (u_{M} - u_{1}) (K - c_{+})\over
J-\alpha}\right]\,, \label{chi_m}\\
u_m &=& u_M + {J-\alpha\over K-c_{+}}\,\chi_m +
B_+\, \left[ e^{(K-c_{+}) \chi_{m}} - 1\right]\,.
\label{u_m}
\end{eqnarray}
If these expressions are inserted in the bias
condition (\ref{bias}), $c_+$, $\chi_m$ and
$u_m$ may be determined as functions of $J$
for a fixed $\phi$. Figures \ref{fig5} and
\ref{fig6} show the phase planes and the
trajectories corresponding to $u(\chi;c_{+})$ and
$u(\chi;c_{-})$, respectively, for a given value
of $J=1.3$. Fig.~\ref{fig3} shows $\chi_m$ and
$u_m$ as functions of $J$.

There are two important approximations to the wavefronts of
Theorem 2, which yield either approximately triangular pulses ($c_+\neq
c_-$, $u_m\gg 1$) or the homoclinic pulse ($c_+=c_-= K$). These
two limiting cases are described in detail in Appendix
\ref{lcases}.

\setcounter{equation}{0}
\section{Pulse Dynamics}
\label{sec:1pulse}

\subsection{One pulse far from the boundaries}
Let us consider a single pulse moving far from the boundaries, as
described in the previous Section. Its height and width are
established by imposing the integral condition (\ref{bias}). The
result will show that the pulse is a tall and narrow moving 
layer which changes size as it moves. At certain
stages of the periodic Gunn oscillation (see below), the pulse
height is $u=O(\epsilon^{-{1\over 2}})$, its width is $\Delta y =
O(\epsilon^{{1\over 2}})$, so that the pulse excess area (in $y$
space units) is $O(1)$. The inner core of the pulse contributes
an order 1 amount to the bias whereas its exponential tails
approaching $u_1(J)$ yield an $O(\epsilon)$ correction to its
excess area. Thus we suggest a decomposition of the solution $u$
into an outer solution valid outside the pulse and boundary layers,
and inner solutions comprising the front(s) and boundary layers.
In this Section, we shall calculate the leading order term in
each asymptotic expansion explicitly. The first correction to these results
can be found in Appendix \ref{P1}.

\subsubsection{Outer solution}
The outer solution is
\begin{eqnarray}
u^{outer} = u^{(0)}(y,s) + \epsilon\, u^{(1)}(y,s)
+ O(\epsilon^2)\label{outer1}\\
J = J^{(0)}(s) + \epsilon\, J^{(1)}(s) + O(\epsilon^2) . \label{outer2}
\end{eqnarray}
Here $u^{(0)} = O(1)$ yields an order 1 contribution to the integral
condition (\ref{bias}), of the same order as that provided by integration of
the excess area of the pulse inner core. $\epsilon\, u^{(1)}$ yields an order
$\epsilon$ contribution to (\ref{bias}), of the same order as that provided
by integration of the excess area of the pulse tails. Inserting this Ansatz
into (\ref{eq2}), we obtain
\begin{eqnarray}
J^{(0)} - g(u^{(0)}) &=& 0,\label{outer3}\\
J^{(1)} - g'(u^{(0)})\, u^{(1)} &=& {\partial u^{(0)}\over\partial s}
+ K\, {\partial u^{(0)}\over\partial y}\,.\label{outer4}
\end{eqnarray}
Solving (\ref{outer3}) and (\ref{outer4}) yields 
\begin{eqnarray}
u^{(0)} = u_1(J^{(0)}) = {J^{(0)}\over\beta}\,,\label{outer5}\\
u^{(1)} = {J^{(1)} - {1\over g'_{1}}\, {dJ^{(0)}\over ds}\over g'_{1}}
 = {1\over\beta}\, \left( J^{(1)} - {1\over \beta}\,
{dJ^{(0)}\over ds}\right)\,.\label{outer6}
\end{eqnarray}
$J^{(0)}$ and $J^{(1)}$ will be found later from the integral constraint.

\subsubsection{Two-term description of the pulse}
The pulse described in the previous Section may be considered a moving inner
layer solution. Its height is much larger than 1 and its width much smaller
than 1 (the precise orders will be determined later). We shall assume
\begin{eqnarray}
u^{inner} \sim P^{(0)}(y,s;\epsilon) + \epsilon\, P^{(1)}(y,s;\epsilon),
\label{inner1}
\end{eqnarray}
where $P^{(0)}$ is the pulse solution of (\ref{fronts}) described in
Theorem 2:
\begin{eqnarray}
P^{(0)}(y,s;\epsilon) = u(x-X_+;c_+)\, \theta[\chi_m - (x-X_+)]
\nonumber\\
+ u(X_+ + 2\chi_m -x;c_+)\, \theta[x-X_+ -\chi_m]. \label{inner2}
\end{eqnarray}
$P^{(0)}(y,s;\epsilon)$ and $P^{(1)}(y,s;\epsilon)$ depend on $\epsilon$ 
in such a way that $$ {P^{(1)}(y,s;\epsilon)\over P^{(0)}(y,s;\epsilon)} = 
O(1),\quad\mbox{as}\quad\quad\epsilon\to 0+ $$
uniformly in $y$, $s$. The inner expansion (\ref{inner1}) is chosen
so that the pulse inner core in $P^{(0)}$ yields an
order 1 contribution to the bias condition whereas its tails together
with the inner core of $\epsilon P^{(1)}$ yield an $O(\epsilon)$
contribution. The latter is of the same order as the contribution of
the outer solution, $\epsilon u^{(1)}$, to the bias.

In Eq.\ (\ref{inner2}), the location of the fronts and their velocities
are:
\begin{eqnarray}
X_+ \sim X_+^{(0)}(t) + \epsilon X_+^{(1)}(t),\nonumber\\
c_+\sim c_+^{(0)} + \epsilon c_+^{(1)} , \label{inner3}
\end{eqnarray}
and similarly for $X_-$ and $c_-$. Inserting (\ref{outer2}) and (\ref{inner1})
- (\ref{inner3}) in (\ref{eq1}) and (\ref{bc+}), we obtain for the trailing
front
\begin{eqnarray}
{\partial^{2} P^{(0)}\over\partial\chi^{2}} -
(K-c_{\pm}^{(0)})\, {\partial
P^{(0)}\over\partial\chi}  + J^{(0)} -
g(P^{(0)}) = 0\,,\label{inner4}\\
P^{(0)}(-\infty;c_{+}^{(0)}) = u^{(0)} =
u_{1},\quad P^{(0)}(0;c_+^{(0)}) =
u_M,\nonumber\\
P^{(0)}(\chi_{m};c_{+}^{(0)}) = u_m,\quad\quad
{\partial P^{(0)}\over\partial
\chi}(\chi_{m};c_{+}^{(0)}) = 0 ,
\label{inner6}
\end{eqnarray}
whose solution is (\ref{inner2}), and 
\begin{eqnarray}
{\partial^{2} P^{(1)}\over\partial\chi^{2}} -
(K-c_{\pm}^{(0)})\, {\partial
P^{(1)}\over\partial\chi} -  g'(P^{(0)}) P^{(1)}
= \nonumber\\
{\partial P^{(0)}\over\partial s}
- J^{(1)} - c_{\pm}^{(1)}\,  {\partial
P^{(0)}\over\partial\chi}\,, \label{inner5}\\
P^{(1)}(-\infty;c_{+}^{(0)}) = u^{(1)},\quad \quad
{\partial P^{(1)}\over\partial\chi}(\chi_{m};
c_{+}^{(0)}) = 0 .  \label{inner7}
\end{eqnarray}
The leading front of the pulse obeys similar expressions. The correction
$P^{(1)}$ is calculated in Appendix \ref{P1}, in which explicit formulas for
piecewise linear $g(u)$ are given.

\subsection{General equation for $J^{(0)}$}
Using Theorem 2, we can calculate the bias
condition for a  single pulse moving far from the
boundaries as
\begin{eqnarray}
\phi = u_1(J^{(0)}) + 2\epsilon\int_0^{\chi_{m}} (P^{(0)} - u_1)\, d\chi
\nonumber\\
+ 2\epsilon\int_{-\infty}^{0} (P^{(0)} - u_1)\,
d\chi  + \epsilon {J^{(1)} - g'_{1}
J_{s}^{(0)}\over g_{1}^{'\, 2}}\nonumber\\ + 2
\epsilon^{2} \int_0^{\chi_{m}} \left(P^{(1)} -
{J^{(1)} - g'_{1}  J_{s}^{(0)}\over g_{1}^{'\,
2}}\right)\, d\chi \nonumber\\
+ \epsilon\int_0^{\infty} [U_L(\xi)- u_1]\, d\xi
\nonumber\\ + \epsilon \int_0^{\infty} [U_R(\xi)-
u_1]\, d\xi + O(\epsilon^2).
\label{bias0}
\end{eqnarray}
Here the bias is the sum of the areas of the regions inside and outside
the moving pulse. The leading order contributions to these areas are
the first two terms on the right side, which are of order 1. They are:
(i) the leading order contribution of the outer solution and (ii)
the leading order contribution of the inner core of the pulse. The other
terms are $O(\epsilon)$ and correspond to:
\begin{itemize}
\item (i) the tails of the pulse
to leading order,
$$ 2\epsilon \int_{-\infty}^0 (P^{(0)} - u_1)\, d\chi = {2
\epsilon(u_{M} - u_{1}) \over\lambda_{+}}\,,$$
\item (ii) the second order contribution to the outer solution,
\item (iii) the second order contribution to the area of the inner core of the
pulse, $2 \epsilon^{2} \int_0^{\chi_{m}} P^{(1)} d\chi = O(\epsilon)$
(we have ignored a much smaller term of order $\epsilon^2 \chi_m$),
\item (iv) the layer at the left boundary, Eqs.~(\ref{eq7}) and (\ref{eq8}):
$$\epsilon \int_0^{\infty} (U_L - u_1)\, d\xi = {2\epsilon\, (\rho -
\beta^{-1})\,J^{(0)}\over \sqrt{K^{2}+4\beta}-K}\,,
$$
\end{itemize}
The area of the injecting (left) boundary layer becomes of order 1
when a new pulse is being shed; otherwise it is of order $\epsilon$, as
indicated above.

If the pulse is far from the boundaries only the first two terms are of
order 1, and we have
\begin{eqnarray}
\phi &=& u_1(J^{(0)}) + 2\epsilon\,\left[ \chi_m (u_M - u_1 - B_+)\right.
\nonumber\\
&+&  \left. {(J^{(0)}-\alpha)\chi_{m}^{2}\over 2(K-c_{+}^{(0)})} + {B_{+}\,
\left( e^{(K-c_{+}^{(0)}) \chi_{m}} -1\right)\over K-c_{+}^{(0)}}
\right] + O(\epsilon) \nonumber\\
&=&  u_1 + 2\epsilon\,\left[ \chi_m (u_M - u_1 - B_+)
+ {(J^{(0)}-\alpha)\chi_{m}^{2}\over 2(K-c_{+}^{(0)})}\right.
\nonumber\\
&+& \left. {B_{+}\lambda_{+} (u_{M} - u_{1})\over J^{(0)}-\alpha}\right]
 + O(\epsilon). \label{bias1}
\end{eqnarray}

\begin{figure}
\begin{center}
\includegraphics[width=8cm]{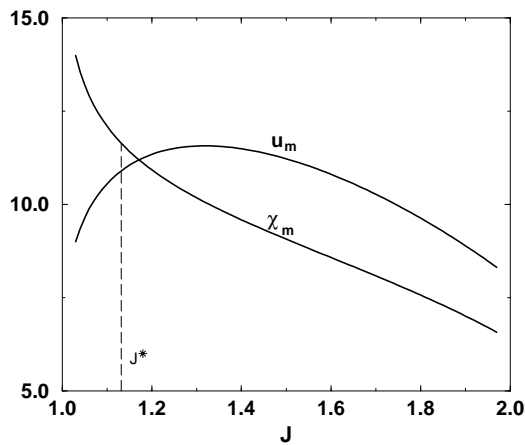}
\caption{ Half-width $\chi_m$ and maximum height $u_m$ of the
single pulse as a function of $J$ for the same parameter values
as in Figures \ref{fig1} to \ref{fig6}.  } \label{fig3}
\end{center}
\end{figure}

\begin{figure}
\begin{center}
\includegraphics[angle=270,width=8cm]{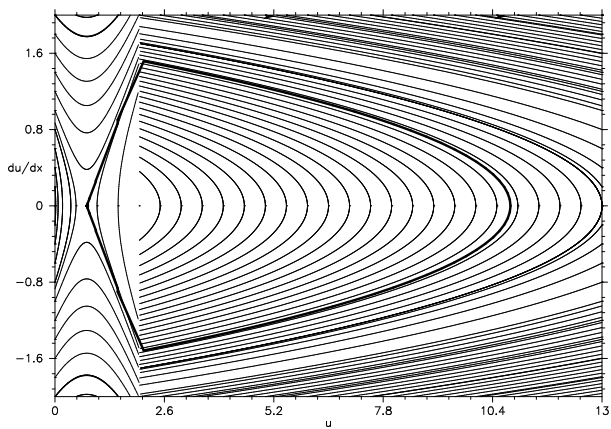}
\caption{  Phase plane $(u,du/d\chi)$ and
homoclinic orbit  for $J^*=1.13182$, $c = 2$.
Other parameter values are as in Figure \ref{fig1}. }
\label{fig7}
\end{center}
\end{figure}

Now we can proceed as sketched in Section \ref{sec:ssm}. Eq.\ (\ref{chi_m})
allows us to obtain $c_+^{(0)}$ as a function of $J^{(0)}$ and $\chi_m$:
\begin{eqnarray}
c_+^{(0)} = \Xi(J^{(0)},\chi_m), \label{cplus}
\end{eqnarray}
for a fixed value of $\phi$. Inserting this function in (\ref{bias1}), we can
determine $\chi_m$ as a function of $J^{(0)}$. Then (\ref{u_m}) yields $u_m$
as a function of $J^{(0)}$. The results are certain functions:
\begin{eqnarray}
\chi_m = {\varphi(J^{(0)})\over 2}\,,\quad u_m =
U(J^{(0)}), \label{Xi1}
\end{eqnarray}
that have been depicted in Fig.~\ref{fig3}. Time differentiation of
$2\chi_m = \varphi(J^{(0)})$ yields Eq.~(\ref{eq_J1}):
\begin{equation}
{dJ^{(0)}\over ds} = 2{c_{+}^{(0)}(J^{(0)},U(J^{(0)})) - K\over
-\varphi'(J^{(0)})},
\label{evolutionJ}
\end{equation}
which describes the time evolution of $J^{(0)}$. Provided that $J^{(0)}(0)>
J^*$, $J^{(0)}$ decreases exponentially fast to $J^*$ such that $c_+^{(0)} =
c_-^{(0)} = K$. The pulse then moves at constant $J^{(0)}=J^*$ and speed $K$,
and it is a homoclinic orbit of the phase plane (\ref{fronts}) with $c=K$, as
shown in Fig.~\ref{fig7}.

\begin{figure}
\begin{center}
\includegraphics[width=10cm]{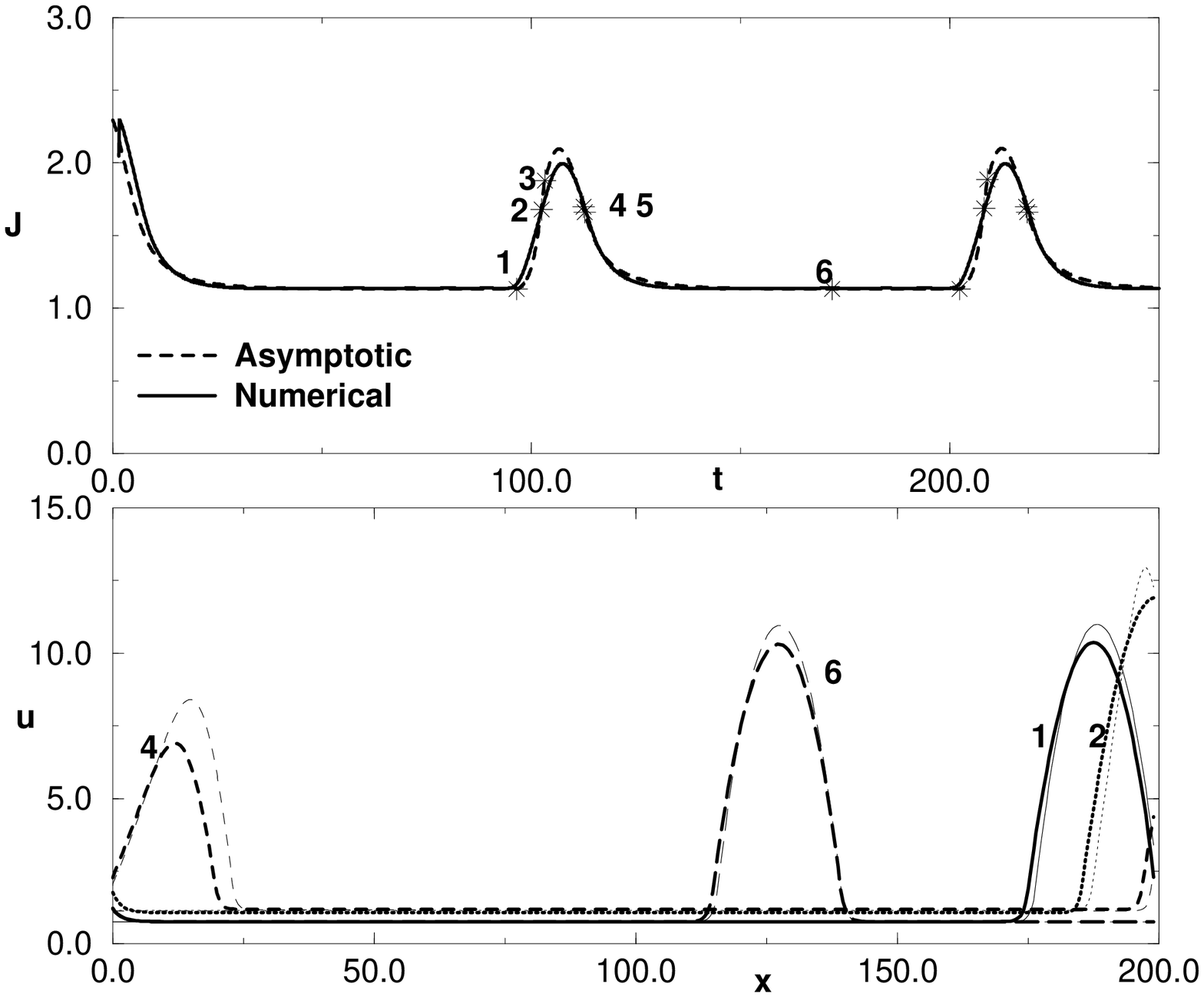}
\caption{Upper: time evolution of $J$. Times marked
correspond to; (1) wave reaches receiving contact, (2) $J$
reaches $J_c$ (3) $\chi_m < \chi_L$ (4) wave completely exits (5)
$J < J_c$ (6) fully developed wave. Lower: profiles of $u(x,t)$ at
the times marked in the upper figure. }
\label{perfiles}
\end{center}
\end{figure}

Figure \ref{perfiles} compares the leading order asymptotic
solution with the numerical solution. The upper part shows the
time evolution of $J(t)$ and the lower part $u(x,t)$ at the times marked in
the upper figure. Notice that initially $J(t)$ decreases exponentially
fast to $J^*$. Consider an instant (point 6) at which the wave is fully
developed and far from the boundaries. We observe that the profile of the
asymptotic solution has a larger height and is slightly thinner than the
numerical solution. Why? We have neglected $O(\epsilon)$ terms (boundary
layers, wave tails, \ldots) when calculating the integral condition with the
leading order asymptotic solution. Then $J^{(0)}(t)$ is slightly larger than
the numerically calculated $J(t)$. The asymptotic profile fully agrees with
the homoclinic solution described in Appendix \ref{lcases} by
(\ref{sol1.homoclinic}), (\ref{sol2.homoclinic}) and (\ref{height.homoclinic})
(also calculated excluding order $\epsilon$ effects).

The previous ideas are correct if we can show
that $J^{(0)}$ evolves  on a slow time scale, say
$\sigma = \sqrt{\epsilon} (t-t_0)$ or $\tau =
\epsilon (t-t_0)$. To see this we should analyze
(\ref{bias1})  and the previous equations with a
little more care. We shall show  that there are
two limiting cases for which the pulse can be
easily calculated and (\ref{eq_J1}) explicitly
obtained to  leading order:
\begin{itemize}
\item $J^{(0)}-J^* = O(1)$, $K-c_+^{(0)} = O(1)$
and $(J^{(0)} -\alpha)\chi_m \gg 1$. The limiting
pulse is a triangular wave formed by pieces of
phase plane trajectories which do not tend
exponentially to infinity as $\chi\to\pm\infty$.
\item $(K-c_+^{(0)}) \ll (J^{(0)}-\alpha)\ll 1$,
$\chi_m \gg 1$.  The limiting pulse is the
homoclinic trajectory with $c=K$.
\end{itemize}
In the first case, $J^{(0)}$ evolves on the slow
time scale $\sigma  = \sqrt{\epsilon} (t-t_0)$.
In the second case, the time over which $J^{(0)}$
varies appreciably is $o(\epsilon^{-1})$. One
period of the Gunn oscillation contains stages
during which these limiting cases are good
approximations to (\ref{eq_J1}). In fact the Gunn
oscillation may be considered as transitions
from one limiting case to the other depending  on
the number of pulses existing at the time. Then
$J^{(0)}$ changes  slowly with time which
justifies our wavefront constructions.

Appendix \ref{app-lim} analyzes the dynamics of pulses
corresponding to the two limiting cases of triangular and
homoclinic pulses. At this point, we may call attention to two
peculiar features of our  results:
\begin{enumerate}
\item Our leading-order approximation for the solution is of order one
outside a pulse, but it is much larger inside it: of order $\epsilon^{-
{1\over 2}}$.
\item The proper time scale for the variation of $J$ is $\epsilon^{
{1\over 2}}t$ if the pulse may be approximated by a triangular wave.
It is slower as the pulse approaches a homoclinic pulse. Both these
approximations are limits of the same equation, (\ref{eq_J1}). Thus
this equation contains more than one asymptotic limit and its time
scale is changing as time changes.
\end{enumerate}

\setcounter{equation}{0}
\section{Pulse dynamics near the boundaries}
\label{sec:dynamics}
In the previous Section, we described the dynamics of a single
pulse far from the boundaries. Basically the pulse approaches
the homoclinic pulse on a slow time scale: at each time on the
scale $t$ the pulse follows adiabatically the instantaneous value
of $J$. $J$ changes on the scale $\sigma=\sqrt{\epsilon} t$ or even more
slowly as described by Eq.~(\ref{eq_J1}), where $\varphi$ is a
function of $J$ and $\epsilon$ ($\varphi(J)$ is of order
$\epsilon^{-{1\over 2}}$ or larger). In this Section we shall
describe what occurs when the pulse reaches the receiving
boundary at $x=L$ and beyond, until a period of the Gunn
oscillation is completed.

\subsection{Pulse disappearing through the receiving boundary}
Let us assume that a single pulse has reached its asymptotic
shape ($J^{(0)}=J^*$) and it advances with speed $K$ until its forefront
reaches $X_-=L$ at time $t_1$. Afterwards, it begins leaving the
sample. As time elapses the wave {\em exits} through the receiving
boundary and therefore the area under the wave decreases. Since the
total area has to satisfy the bias condition, this loss of area has
to be compensated by a corresponding increase in $u_1$ so that
equation (\ref{bias}) still holds.

Let us denote by $u_L$ the value of the pulse inner solution
$P^{(0)}(y,s)$ at the receiving boundary. $u_L$ is obtained from
equation (\ref{>0}) for $u(x-X_-^{(0)};c_-^{(0)}) = u(X_-^{(0)} - x;
c_+^{(0)}) = u(2\chi_m - \chi;c_+^{(0)})$ when $\chi - 2\chi_m =
L-X_-^{(0)}$. The corresponding argument of $u(X_-^{(0)} - x;
c_+^{(0)})$ is $\chi_L = X_-^{(0)} - L >0$. When $0<\chi_L < \chi_m$,
the bias condition (\ref{bias}) yields
\begin{eqnarray}
\phi = u_1 +\Phi^{(0)}(J,c_+^{(0)},\chi_m,\chi_L)
+ \epsilon\,\Phi^{(1)}(J,c_+^{(0)},\chi_m,
\chi_L,J^{(1)})  + O(\epsilon^2),  \label{bias4}\\
{\Phi^{(0)}(J,c_+^{(0)},\chi_m,\chi_L)\over
\epsilon} = 2 \int_0^{\chi_{m}}
[u(\chi;c_+^{(0)})-u_1]\, d\chi  
-\int_0^{\chi_{L}} [u(\chi;c_+^{(0)}) -u_1]\,
d\chi \nonumber\\ 
= (2\chi_m - \chi_L)\, (u_M - u_1 - B_+) 
+{(J-\alpha)\over (K-c_{+}^{(0)})}\,
\left(\chi_{m}^{2}  - {\chi_{L}^{2}\over
2}\right)\nonumber\\
+ {B_{+}\,\left( 2 e^{(K-c_{+}^{(0)})
\chi_{m}} -  e^{(K-c_{+}^{(0)}) \chi_{L}} -
1\right)\over K-c_{+}^{(0)}} \, , \label{Phi}\\
\Phi^{(1)}(J,c_+^{(0)},\chi_m,\chi_L,J^{(1)}) = \int_{-\infty}^{0}
(P^{(0)}-u_1) d\chi + {J^{(1)} - g'_{1} J_{s}^{(0)}\over g_{1}^{'\, 2}}
\nonumber\\
+ 2 \epsilon \int_0^{\chi_{m}} P^{(1)}\, d\chi 
- \epsilon \int_0^{\chi_{L}} P^{(1)}\, d\chi
+ \epsilon\int_0^{\infty} [U_L(\xi)- u_1]\, d\xi ,\label{Phi1}
\end{eqnarray}
instead of (\ref{bias0}). In this equation $\Phi^{(0)}$ and $\Phi^{(1)}$
are of order 1 because the integrations of $P^{(0)}$ and $P^{(1)}$ over
the inner core of the pulse are of order $\epsilon^{-1}$. Eq.~(\ref{bias4})
yields
\begin{eqnarray}
\phi = u_1 + \Phi^{(0)}(J^{(0)},c_+^{(0)},\chi_m,\chi_L), \label{phi0}\\
\Phi^{(1)}(J^{(0)},c_+^{(0)},\chi_L,\chi_m,J^{(1)}) = 0.  \label{phi1}
\end{eqnarray}

We shall now find the evolution equation for $J^{(0)}$ by a procedure similar
to that used to find (\ref{evolutionJ}). The right side of (\ref{phi0})
depends on $J^{(0)}$, $c_+^{(0)}$, $\chi_m$ and $\chi_L$. $\chi_m$ is a
function of $J^{(0)}$ and $c_+^{(0)}$ given by (\ref{chi_m}). As wavefront
velocity $c_+^{(0)}$, we shall use the function of $J^{(0)}$ (for a fixed
value of $\phi$) that was determined at the end of Section \ref{sec:univ}.
Then the right side of (\ref{phi0}) is a function of $J^{(0)}$ and $\chi_L$
only (for fixed $\phi$): 
\begin{eqnarray}
\phi &=& {\cal B}(J^{(0)},\chi_L) \equiv u_1(J^{(0)}) \nonumber\\ 
&+&  \Phi^{(0)}(J^{(0)},c_+^{(0)}(J^{(0)}),\chi_m(J^{(0)},
c_+^{(0)}(J^{(0)})),\chi_L ).  \label{calB}
\end{eqnarray} 
$\chi_L$ can be explicitly calculated from
\begin{equation}
{d\chi_{L}\over dt} = c_-^{(0)} = 2K - c_+^{(0)} ,\quad \chi_L(t_1) = 0,
\label{c-}
\end{equation}
where $c_+^{(0)}$ is our known function of $J^{(0)}$. We can obtain a closed
system of equations for $\chi_L$ and $J^{(0)}$ by differentiating
(\ref{calB}) with respect to time and then using (\ref{c-}). The result is
\begin{eqnarray}
{\partial {\cal B}\over\partial J^{(0)}}\, {dJ^{(0)}\over dt} \sim
\epsilon \, (u_L - u_1)\,
(2K-c_+^{(0)})\,. \label{bias5}
\end{eqnarray}
Here we have used that (\ref{Phi}) and (\ref{calB}) imply $\partialÊ{\cal B}/
\partial \chi_L = - (u_L -  u_1)$. 

The time evolution of $J^{(0)}$ is found by solving this equation while the
wave disappears through the receiving contact. Having found the solution to
leading order, (\ref{phi1}) yields the correction $J^{(1)}$. Numerical
solution of (\ref{bias4}) to (\ref{bias5}) shows that $J^{(0)}$ increases
with time, as shown in the region between times 1 and 2 in figure
\ref{perfiles}. Notice that this increase agrees with the numerical solution
of (\ref{eq1}-\ref{ic}).

Depending on the bias $\phi$, one of the following two events may
happen first:
\begin{itemize}
\item (i) $J^{(0)}$ reaches $J_c$, or
\item (ii) $\chi_L = \chi_m$.
\end{itemize}
In both cases the stage described by the previous equations ends.
In case (i), a new wave is created at $x=0$, whereas in case (ii)
(\ref{bias4}) should be changed to
\begin{eqnarray}
\phi = u_1 + \Psi^{(0)}(J^{(0)},c_+^{(0)},\chi_L) %\nonumber\\
+ \epsilon\, \Psi^{(1)}(J^{(0)},c_+^{(0)},\chi_L,J^{(1)}) + O(\epsilon^2),
\label{bias6}\\
\Psi^{(0)}(J^{(0)},c_+^{(0)},\chi_L) = \epsilon\,\int_0^{\chi_{L}}
[u(\chi;c_+^{(0)})-u_1]\, d\chi \nonumber\\
= \epsilon\,\left[{(J^{(0)}-\alpha)\chi_{L}^{2}\over 2(K-c_{+}^{(0)})}
+ \chi_L (u_M - u_1 - B_+)
+ {B_{+}\, \left( e^{(K-c_{+}^{(0)}) \chi_{L}} -1\right)\over
K-c_{+}^{(0)}}\right]\,, \label{bias7}\\
\Psi^{(1)}(J,c_+^{(0)},\chi_L,J^{(1)}) = \int_{-\infty}^{0}
[u(\chi;c_+^{(0)})-u_1] d\chi
+ {J^{(1)} - g'_{1} J_{s}^{(0)}\over g_{1}^{'\, 2}}\nonumber\\
+ \epsilon \int_0^{\chi_{L}} P^{(1)}\, d\chi 
+ \epsilon\int_0^{\infty} [U_L(\xi)- u_1]\, d\xi ,\label{Psi1}
\end{eqnarray}
where now $\chi_L = L-X_+^{(0)} >0$ and
\begin{equation}
{d\chi_{L}\over dt} = - c_+^{(0)} ,\quad \chi_L(t_2) = \chi_m(t_2) .
\label{c+}
\end{equation}
Here $t_2$ is the time at which the maximum of the pulse reaches $x=L$
(equivalently $\chi_L=\chi_m$ in the previous stage). Similar arguments to
those used in the previous stage lead to,
\begin{eqnarray}
\left( {1\over\beta} + {\partial\tilde{\Psi}^{(0)}\over\partial
J^{(0)}}\right)\, {dJ^{(0)}\over dt} \sim \epsilon \, (u_L -
u_1)\, c_+^{(0)}\,,
\end{eqnarray}
where $\tilde{\Psi}^{(0)}(J^{(0)},\chi_L) = \Psi^{(0)}(J^{(0)},c_+^{(0)}
(J^{(0)}),\chi_L)$.

This stage lasts until $J^{(0)}$ reaches $J_c$ and a new pulse is shed.
If the bias is small enough, the solution of these equations may be
such that $J^{(0)}$ never reaches $J_c$ and it eventually decreases
with time. In such case, the pulse exits and leaves a stable
stationary state in its wake after $\chi_L = 0$. Notice that setting
$\chi_L = \chi_m = 0$ in (\ref{chi_m}) implies a front
velocity (\ref{c+.limit}). Thus the velocity of the disappearing
wavefront approaches that of the triangular wave, although the
shapes of the respective wavefronts may differ greatly. As in the
previous stage, we find $J^{(1)}$ by solving the equation
$\Psi^{(1)}(J^{(0)},c_+^{(0)},\chi_L,J^{(1)}) = 0$.

\subsection{Pulse shedding at the injecting boundary}
If $J(t)$ reaches $J_c$ at $t=t_2$, the boundary layer
becomes unstable and a new pulse starts being shed at the
injecting boundary. The boundary layer profile, $U(x,t)$ solves
the following semiinfinite integrodifferential problem:
\begin{eqnarray}
{\partial U\over\partial\sigma} + K{\partial U\over\partial x} -
{\partial^{2}U\over\partial x^{2}} + g(U) = J(\sigma),\label{U1}\\
U(0,\sigma) = \rho\, J, \label{U3}
\end{eqnarray}
whose solution exhibits an explosion-type behavior and a rapid
growth of the area enclosed by the boundary layer which can no
longer be neglected. This increase in area is to be compared with
the area released by the disappearing pulse at $x=L$. Initially,
the area released  is larger and this net area loss has to be
compensated by an equal increase in the area of the outer
solution, so that $J$ continues to increase although at a slower
rate. After a short time, the growth of the boundary layer is
larger than the area released, so that $J$ reaches a maximum and
starts decreasing.

The structure of the injecting boundary layer when $J>J_c$ is as
follows:
\begin{enumerate}
\item $u=u(x;J,U_m)$ is quasi-stationary for $0<x<X_m$ such that
(\ref{eq5}) holds with $u(0;J,U_m) = \rho J$, $u(X_m;J,U_m) = U_m$,
with $\partial u(X_m;J,U_m)/\partial x = 0$. The boundary layer
solution is the trajectory of the phase plane corresponding to
(\ref{eq5}) which leaves the vertical line $u=\rho J$ at $x=0$
and intersects the $u$ axis at $u=U_m$. Notice that this
trajectory is uniquely determined by giving $J$ and $U_m$.
\item For $x>X_m$, the boundary layer is a wavefront of type
$u(\chi;c_-)$ moving at speed $C_-$. This speed and the forefront
are uniquely determined  by $J$ and $U_m$: $u = u(\xi;C_+)$,
$\xi = Xn_- - x$, with $C_+ = 2 K - C_-$, $u(-\infty;C_+) = u_1(J)$,
$u(0;C_+) = u_M$, $u(\xi_m;C_+) = U_m$, $\partial u(\xi_m;C_+)/
\partial\xi = 0$.
\end{enumerate}

The bias condition (including the injecting boundary layer) is
\begin{eqnarray}
\phi = u_1 + \tilde{\Psi}^{(0)}(J,\chi_m,\chi_L) + \epsilon\, A^{(0)}(t)
%(J,\xi_m)
+ O(\epsilon),\label{bias8}\\
A^{(0)}
%(J,\xi_m)
= \int_0^{\infty} [U(\xi)-U_{\infty}]\, d\xi ,
\label{bias9}
\end{eqnarray}
provided that the maximum of the old pulse has leaved the sample.
Time differentiation of (\ref{bias8}) yields
\begin{eqnarray}
\left( {1\over\beta} +  {\partial \tilde{\Psi}^{(0)}\over\partial J^{(0)}}
\right)\, {dJ^{(0)}\over dt} \sim - \epsilon {d A^{(0)}\over dt}  + \epsilon
\, (u_L - u_1)\, c_+^{(0)} . \label{bias11}
\end{eqnarray}
Equations (\ref{c+}), (\ref{bias8}),
and (\ref{bias11}), together with (\ref{Phi}), (\ref{bias7}) and
(\ref{Xi1}), constitute a closed system of equations for the
unknowns $J$, $\chi_L$, and $\chi_m$. During this stage, $J$ initially increases 
and then decreases
until it either reaches again $J_c$, or the old wave completely
disappears. In the latter case, the evolution of $J(t)$ is still
given by equation (\ref{bias11}) without the last term which
represents the area lost by the disappearing wave.
The evolution during this stage can be observed between points 3
and 4 in Figure \ref{perfiles}.

\begin{figure}
\begin{center}
\includegraphics[width=8cm]{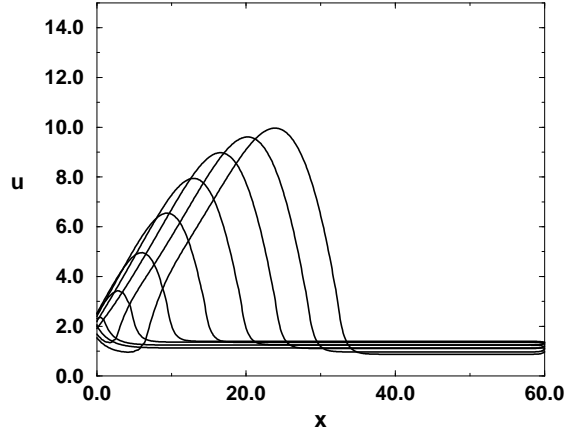}
\caption{Boundary layer profile evolution during the shedding stage. }
\label{figbl}
\end{center}
\end{figure}

Also, Figure \ref{figbl} shows in detail the evolution of the boundary layer
profile from the time that $J > J_c$. Initially, $u$ grows at the injecting
boundary because $J$ increases. Furthermore, the slope $\partial u(0,t)/
\partial x$ increases and it becomes positive at a certain time. Then a
wave-like structure is created. The leading front of this wave moves away
from the boundary while the back-front is attached to it. The wave continues
its growth until the time when $J$ becomes again smaller than $J_c$. Then,
the slope at the boundary becomes negative, the wave dettaches and moves away
from the boundary as a solitary wave. At that time, the quasistationary part
of the boundary layer which joins $x=0$ to the maximum at $x=X_m$ becomes the
backfront of a detached pulse. Then, we again have the same equation
(\ref{eq_J1}) describing the first stage, and a cycle of the Gunn oscillation
has been completed. The time evolution during this stage can be observed in
Figure \ref{perfiles} for times larger than $t_5$.

\setcounter{equation}{0}
\section{Conclusions}
\label{sec:conclusions}
In this paper we have asymptotically described one period of
a Gunn-type oscillation in a simple model. The nonlinearity
of the model is such that at most two constant solutions
are possible for each value of $J$ (the current-like unknown).
The model consists of a parabolic equation for the field-like
unknown, $u(x,t)$, and an integral constraint (bias condition)
which determines $J(t)$. Appropriate boundary and initial conditions
are imposed. The key new idea of our analysis is that a pulse
that changes shape as it advances may be constructed by fixing
only two parameters: $J$ and the pulse maximum, $u_m$. If the pulse 
width is small compared to the sample length, $L$, $J$ and $u_m$ 
change on a slow time scale. The trailing front of the pulse is part of 
a separatrix joining a saddle
point to $(u_m,0)$ with $du/d\chi>0$ on the $(u,du/d\chi)$ phase
plane. The initial and final points determine the backfront
speed $c_+$ as a function of $J$ and $u_m$. Similarly the
forefront of a pulse is constructed and its speed $c_-$ determined.
Then equations for $J$ and $u_m$ are obtained by time-differentiation
of the bias condition and of the pulse width. The time derivative
of the later is $c_- - c_+$ and $J$ tends towards a fixed value
corresponding to a rigidly moving pulse with $c_- = c_+$. 
Other stages of a Gunn oscillation including wave creation and
annihilation at the boundaries are analyzed by similar methods.
Our theory compares well with direct numerical simulations.

We have found an asymptotic theory of the `Gunn effect' in a simple
piecewise linear model, whose main step is a construction of pulses and
a derivation of an equation for the `current'. An analysis of the stability of 
these solutions is an open problem, although there is some work on this problem
in the related Kroemer model \cite{BHV94,vol69}. That the profile of pulses can 
become oscillatory for appropriate parameter values has been shown by Sun et al 
for some reaction-diffusion models \cite{STWW}. We expect that the present method 
yield results independent of the model equations within a class thereof displaying the
Gunn instability \cite{BC96,samu95,shaw79,vol69,wac02}. 
Studies of other systems that can be understood by the dynamics of pulses 
are in progress.

\renewcommand{\theequation}{A.\arabic{equation}}
\setcounter{equation}{0}
\appendix
\section{Limiting cases}
\label{lcases}
\subsection{Triangular pulses}
The bias condition (\ref{bias}) determines the orders of
magnitude of $\chi_m$ and $u_m$ in terms of the small parameter
$\epsilon$. Let us assume that $(K-c_+)$ and $(J-\alpha)$ are
$O(1)$, whereas $\chi_m \gg 1$. Then (\ref{B+}), (\ref{chi_m})
and (\ref{u_m}) imply that
\begin{eqnarray}
u_m &=& u_M + {J-\alpha\over K-c_{+}}\,\chi_m - {\lambda_{+}
(u_{M}- u_{1})\over K-c_{+}} \nonumber\\
&\sim& {J-\alpha\over K-c_{+}}\,\chi_m .\label{max1}
\end{eqnarray}
The wavefront $u(\chi;c_+)$ is given by (\ref{<0}) for $\chi <0$
and
\begin{eqnarray}
u &=& u_m + {J-\alpha\over K-c_{+}}\, (\chi - \chi_m) + B_+ \, 
\left[ e^{(K-c_{+})\chi} -
e^{(K-c_{+})\chi_{m}}\right]
\label{eq_u}\\
&\sim& %\sim
{J-\alpha\over K-c_{+}}\,\chi ,\label{max2}
\end{eqnarray}
for $\chi >0$, where (\ref{max1}) has been used. The bias
condition (\ref{bias}) then yields
\begin{eqnarray}
{\phi- u_{1}\over\epsilon} \sim {J-\alpha\over K-c_{+}}\,\chi_m^2
,
\label{bias2}\\
\chi_m \sim \sqrt{{(\phi- u_{1})\, (K-c_{+})\over\epsilon\,
(J-\alpha)}} \,. \label{bias3}
\end{eqnarray}
These equations imply that $\chi_m$ and $u_m$ are $O(\epsilon^{
-{1\over 2}})$, while the proper time scale over which $J$ varies
is $t=O(\epsilon^{-{1\over 2}})$.

We can obtain (\ref{max1}) - (\ref{bias3}) directly from the
Eq.\ (\ref{fronts}) and the bias condition. If $\chi_m\gg 1$, so is
$u_m$. Then we shall select uniquely the wavefront solution of 
Eq.\ (\ref{fronts}), $u(\chi;c_+)$, so that it satisfies $u(-\infty;c_+) =
u_1(J)$ and it does not grow exponentially as $\chi\to +\infty$.
Similarly, $u(\chi;c_-) = u(-\chi;c_+)$ does not grow
exponentially as $\chi \to -\infty$, and it satisfies
$u(+\infty;c_-) = u_1(J)$. As in Theorem 1, we still have $c_+ + c_- = 2K$.

\begin{figure}
\begin{center}
\includegraphics[angle=270,width=8cm]{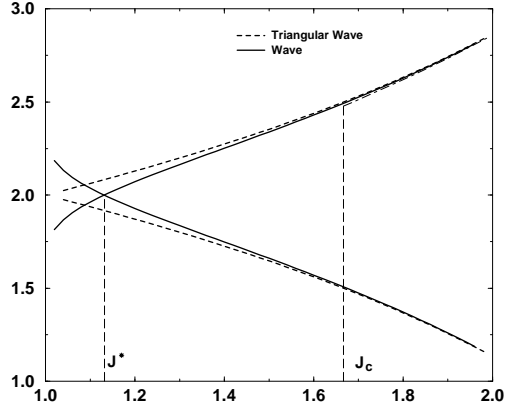}
\caption{Velocities of the backfront, $c_+$, and of the forefront,
$c_-$, as functions of $J$ for $\phi = 1.6$ and the same parameter
values as in Figure \ref{fig1}. We have also shown the corresponding
approximate values $c_{\pm}$ obtained for the triangular pulse
described in the text.
}
\label{fig2}
\end{center}
\end{figure}

$u(\chi;c_+)$ satisfies Eq.~(\ref{<0}) and
\begin{equation}
u(\chi;c_+) = u_M + {J - \alpha\over K- c_{+}}\,\chi,
\quad\mbox{if}\quad \chi >0. \label{lin.growth}
\end{equation}
Continuity of $du/d\chi$ at $\chi = 0$ directly yields
\begin{eqnarray}
\left( u_M - {J\over\beta}\right)\, \lambda_+ = {J - \alpha\over
K- c_{+}}\,,\nonumber
\end{eqnarray}
which in turn implies
\begin{eqnarray}
c_{+} = K - \frac{J - \alpha}{\sqrt{\beta\, \left( u_{M} - u_{1}
 \right)\, \left( u_{M} - {\alpha\over\beta}
\right)} }\, > 0.    \label{c+.limit}
\end{eqnarray}
Figure \ref{fig2} compares the actual values of $c_{\pm}$ with the
approximation (\ref{c+.limit}). Notice that both lines  are
resonably close for $J$ sufficiently higher than $J^*$.

\begin{figure}
\begin{center}
\includegraphics[width=8cm]{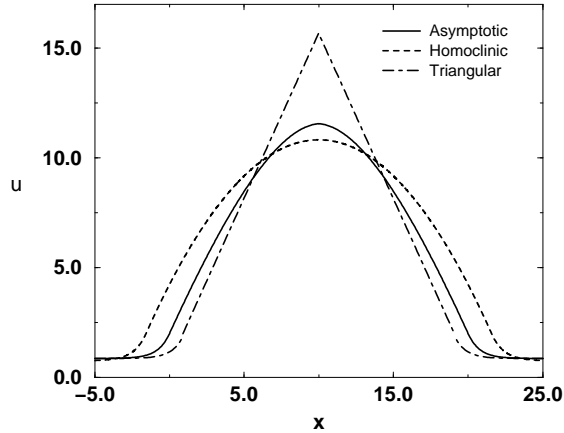}
\caption{ Shape of the pulse for $J=1.3$
and the parameter values of Figure \ref{fig1}. Also shown
are the corresponding triangular and homoclinic pulses.  }
\label{fig4}
\end{center}
\end{figure}

We can now form a pulse by joining the backfront $u(\chi;c_+)$,
$\chi = x-X_+ < \chi_m$ to the forefront $u(2\chi_m-\chi;c_+)$.
This pulse asymptotically approaches an isosceles triangle of
basis $(X_- - X_+) = 2\chi_m$ and height approximately given by
\begin{eqnarray}
u_m = u(\chi_m;c_+) = u\left({X_{-} - X_{+}\over 2};c_+\right)
\nonumber\\
\sim {(J - \alpha)\, (X_{-} - X_{+})\over 2 (K- c_{+})}\,.
\label{height}
\end{eqnarray}
See Fig.~\ref{fig4}, which compares the triangular pulse to the
real pulse and the homoclinic pulse for the same values of $J$ and
$\phi$. Here we have used (\ref{lin.growth}) and assumed that
$\chi_m$ (the location of the maximum, equal to the pulse
half-width) is very large. To be precise, we assume that $(K-c_+)
= O(1)$ and that $(J-\alpha)\,\chi_m \gg 1$.

Notice that the way we have constructed $u(\chi;c_+)$ is
immediately applicable to a general smooth nonlinearity $g(u)$ of
the same type. We have thus the following result:

\noindent {\bf Result}. {\em Let the characteristic curve $g(u)$
be an odd function of $u$ with a positive local maximum after
which it monotonically decays to a positive constant as $u\to
+\infty$. The approximate backfront $u(\chi;c_+)$ is the unique
solution of Eq.~(\ref{fronts}) which, for an appropriate choice
of the velocity $c_+$, satisfies $u(-\infty;c_+) = u_1(J)$ and it
does not grow exponentially as $\chi\to + \infty$.}

\subsection{The homoclinic pulse}
\begin{figure}
\begin{center}
\includegraphics[width=8cm]{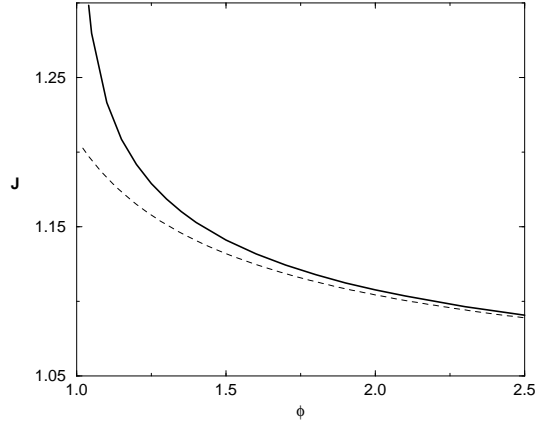}
\caption{ Value $J=J^*$ for the homoclinic pulse as a
function of the bias $\phi$. The dashed curve corresponds
to the approximation (\ref{J.homoclinic}). }
\label{fig8}
\end{center}
\end{figure}

If $c_+=c_-=K$, the pulse is a homoclinic orbit of the phase
plane (\ref{fronts}):
\begin{eqnarray}
{\partial^{2} u\over\partial \zeta^{2}} + J -  g(u) = 0\,,
\label{eq.homoclinic}
\end{eqnarray}
with $u(\pm\infty) = u_1(J) = J/\beta$. See Fig.~\ref{fig7}. Here
$\zeta \equiv x - X_0$, and $X_0 = (X_+ + X_-)/2$ is the location
of the maximum of the pulse, $u_m$. The solution is
\begin{eqnarray}
u(\zeta) = {J\over \beta} + \left( u_{M} - {J\over\beta} \right)\,
e^{- \sqrt{\beta} (|\zeta| - \zeta_{0})} \nonumber\\
\mbox{for}\quad |\zeta| > \zeta_0,\label{sol1.homoclinic}\\
u(\zeta) = u_m - {J-\alpha\over 2}\, \zeta^{2}\,
\quad\mbox{for}\quad
|\zeta| < \zeta_0, \label{sol2.homoclinic}\\
u_m = u_M + {J-\alpha\over 2}\,
\zeta_0^{2}\,.\label{height.homoclinic}
\end{eqnarray}
$\zeta_0$ is determined by imposing continuity of $du/d\zeta$ at
$\zeta =\pm \zeta_0$:
\begin{eqnarray}
\zeta_0 = {\sqrt{\beta} \left( u_{M} - {J\over\beta}\right) \over
J-\alpha}\,.\label{zeta_0.homo}
\end{eqnarray}
Now we may find $J= J^*$ from the bias condition. The result is
\begin{eqnarray}
J^* = \alpha + \beta^{{3\over 4}} \sqrt{ {2 \epsilon \left( u_{M}
- {\alpha\over\beta} \right)^{3}\over 3 \left(\phi - {\alpha\over
\beta} \right)} } + O(\epsilon),
\label{J.homoclinic}\\
\zeta_0 = \beta^{- {1\over 4}} \sqrt{ {3 \left(\phi - {\alpha
\over\beta}\right)\over 2\epsilon\, \left( u_{M} -
{\alpha\over\beta}\right) }} + O(1),
\label{width.homoclinic}\\
u_m = {\beta^{{1\over 4}}\over 2}\, \sqrt{ {3\over 2\epsilon}\,
\left(\phi - {\alpha\over\beta}\right)\, \left( u_{M} -
{\alpha\over\beta}\right) } + O(1). \label{u_+.homoclinic}
\end{eqnarray}
Fig.~\ref{fig8} compares $J^*$ to the approximation
(\ref{J.homoclinic}). Notice that a simple phase plane argument
indicates that $J^*$ obeys the following equal-area rule:
\begin{eqnarray}
J^* = \frac{1}{u_{m} - u_{1}}\,\int_{u_{1}}^{u_{m}} g(u)\, du\,,
\label{ear}
\end{eqnarray}
where $u_m$ is given by Eq.~(\ref{u_+.homoclinic}).
Fig.~\ref{fig4} compares the actual pulse, the homoclinic pulse
with $c=K$ and the triangular wave for the same values of $J$ and
$\phi$.

\renewcommand{\theequation}{B.\arabic{equation}}
\setcounter{equation}{0}
\section{Explicit calculation of $P^{(1)}$ and $c^{(1)}_+$}
\label{P1}
First of all, $P^{(1)}= P^{(0)}_{\chi}\equiv
\partial P^{(0)}/\partial\chi$ is a  solution of
the homogeneous problem (\ref{inner5}) (with zero
right hand  side) and (\ref{inner7}). The
solvability conditions for the  nonhomogeneous
problem yield
\begin{eqnarray}
\left[P^{(0)}_{\chi}\, P^{(1)}_{\chi} - P^{(0)}_{
\chi\chi}\, P^{(1)} \right]_{\chi=\chi_{m}}\,
e^{-(K-c_{+}^{(0)})\chi_{m}} \nonumber\\
=\int_{-\infty}^{\chi_{m}} P^{(0)}_{\chi}
\left[P^{(0)}_{s} - J^{(1)} - c_{+}^{(1)}
P^{(0)}_{\chi}\right]\,  e^{-(K-c_{+}^{(0)})\chi}
d\chi,    \label{inner8}\\
c_{+}^{(1)} + c_{-}^{(1)} = 0. \label{inner9}
\end{eqnarray}
Let us define
\begin{eqnarray}
P^{(1)}\equiv p(\chi;c_+^{(0)})\, {\partial P^{(0)}\over\partial\chi}\,.
\label{inner10}
\end{eqnarray}
Inserting (\ref{inner10}) in (\ref{inner5}), we
obtain an equation which can be solved by two
quadratures. We find
\begin{eqnarray}
{\partial p\over\partial\chi} = {e^{(K-c
_{+}^{(0)})\,\chi}\over  P^{(0)\, 2}_{\chi}}\,
\int_{-\infty}^{\chi} e^{-(K-c_{+}^{(0)})\,\chi}
P^{(0)}_{\chi}\,\left[P^{(0)}_{s}
- J^{(1)}- c_{+}^{(1)} P^{(0)}_{\chi}
\right]\, d\chi. \label{inner11}
\end{eqnarray}
Here subscripts indicate partial derivatives with respect to the
corresponding variable. It is easy to check that this expression
satisfies (\ref{inner8}). Further integration yields (for $\chi<0$)
\begin{eqnarray}
&& p = {(\beta\, J^{(1)} - J_{s}^{(0)})\, e^{-\lambda_{+}\chi}
\over\beta^{2}\lambda_{+}(u_{M} - u_{1})}
+ {\lambda_{+s}\chi^{2}\over 2\lambda_{+}\sqrt{(K-c_{+}^{(0)})^{2}+4\beta}}
\nonumber\\
&& -{\chi\over \sqrt{(K-c_{+}^{(0)})^{2}+4\beta}}\left[{J_{s}^{(0)} \over
\beta\lambda_{+}(u_{M}-u_{1})} 
+ {\lambda_{+s}\over \lambda_{+}
\sqrt{(K-c_{+}^{(0)})^{2}+4\beta}} + c_{+}^{(1)}\right] + q\,,
\label{inner12}
\end{eqnarray}
where $q$ is a constant. For $\chi>0$, $P^{(0)}_{\chi\chi} =
B_+ (K-c_{+}^{(0)})^{2} e^{(K-c_{+}^{(0)})\,\chi}$ and we can
simplify the expression for $p$ by integrating by parts.
The result is
\begin{eqnarray}
p = {1\over B_{+} (K-c_{+}^{(0)})^{2} P^{(0)}_{\chi}}\, \left\{
P^{(0)}_{\chi}\int_{0}^{\chi} e^{-(K-c_{+}^{(0)})\,\chi}
\right.\nonumber\\
\left.\left[P^{(0)}_{s} - J^{(1)}- c_{+}^{(1)} P^{(0)}_{\chi}
\right]\, d\chi - \int_{-\infty}^{\chi} e^{-(K-c_{+}^{(0)})\,\chi}
\right.\nonumber\\
\left. P^{(0)}_{\chi}\,\left[P^{(0)}_{s} - J^{(1)}- c_{+}^{(1)} P^{(0)}_{\chi}
\right]\, d\chi\right\} + Q. \label{inner13}
\end{eqnarray}
Continuity of $p$ at $\chi=0$ yields
\begin{eqnarray}
q = Q - {J^{(1)} - {J_{s}^{(0)}\over\beta } + {\beta I_{0}\over B_{+}
(K-c_{+}^{(0)})^{2}}\over \lambda_{+}\beta (u_{M}-u_{1})}
\label{inner14}
\end{eqnarray}
where $I_0$ is the following integral:
\begin{eqnarray*}
I_0 \equiv \int_{-\infty}^{0} e^{-(K-c_{+}^{(0)})\,\chi}
P^{(0)}_{\chi}\,\left[P^{(0)}_{s} - J^{(1)}- c_{+}^{(1)} P^{(0)}_{\chi}
\right]\, d\chi
\end{eqnarray*}
whose value can be computed as,
\begin{eqnarray}
I_0 \, = \, I_{00} \, + \, I_{0J} \, J^{(1)} \, + \, I_{0c} \, c_{+}^{(1)}
\label{inner14bis}
\end{eqnarray}
where
\begin{eqnarray*}
I_{00} &=&
{\lambda_{+}^{2}(u_{M}-u_{1})\over\sqrt{(K-c_{+}^{(0)})^{2}+4\beta}}\,
\left[ \frac{\lambda_{+}J_{s}^{(0)}}{\beta^{2}} \, + \,
\frac{(u_M-u_1)\,  c_{+s}}{(K-c_{+}^{(0)})^{2} + 4 \beta} \right]  \\
I_{0J} &=& - \frac{\lambda_{+}^{2}(u_{M}-u_{1})}{\beta} \\
I_{0c} &=& - \frac{\lambda_{+}^{2}(u_{M}-u_{1})^{2}}{
\sqrt{(K-c_{+}^{(0)})^{2}+4\beta}}
\end{eqnarray*}
Continuity of $P^{(1)}_\chi$ at $\chi=0$ implies
\begin{eqnarray}
P^{(0)}_{\chi}(0)\, \left[ p_{\chi}(0+) - p_{\chi}(0-)\right] =
\quad\quad \nonumber\\
- p(0)\, \left[P^{(0)}_{\chi\chi}(0+) - P^{(0)}_{\chi\chi}(0-)\right]\, .
\label{in1}
\end{eqnarray}
The second argument $c_+^{(0)}$ has been omitted in all the functions in
this formula. The jump discontinuity of the second derivative
$P^{(0)}_{\chi\chi}$ at $\chi = 0$ implies that $p_{\chi}$ has also a jump
discontinuity at $\chi = 0$. Substituting (\ref{inner12}) and (\ref{inner13})
in (\ref{in1}), we obtain
\begin{eqnarray}
Q B_+ (K-c_{+}^{(0)})^{2} = q \lambda_{+}^{2}(u_{M}-u_{1})
- {1\over\sqrt{(K-c_{+}^{(0)})^{2}+4\beta}} \nonumber\\
\times \left[{J_{s}^{(0)}\over\beta} + (u_{M}-u_{1}) \lambda_{+} c_+^{(1)}
+ {\lambda_{+\, s} (u_{M}-u_{1})\over\sqrt{(K-c_{+}^{(0)})^{2}
+4\beta} }\right]\,.\label{in2}
\end{eqnarray}
We can obtain $Q$ from (\ref{inner14}), (\ref{inner14bis}) and (\ref{in2}).
After some algebra, the result is
\begin{eqnarray}
Q \, = \,  Q_{0} \, + \, Q_{J} \, J^{(1)} \, + \, Q_{c} \, c_{+}^{(1)}
\label{in3}
\end{eqnarray}
with
\begin{eqnarray*}
\upsilon \, Q_0 =
{J_{s}^{(0)}\over\beta} \, \left[
\frac{1}{\sqrt{(K-c_{+}^{(0)})^{2} +4\beta}} \, - \,
\frac{\lambda_{+}}{\beta} \right] \, + \\
+ \frac{\lambda_{+} \, I_{00}}{B_{+} (K-c_{+}^{(0)})^{2}} \, - \,
\frac{\lambda_{+} \, (u_{M}-u_{1}) \, c_{+\, s}^{(0)}}
{[(K-c_{+}^{(0)})^{2} +4\beta]^{3/2}} \\
\upsilon \, Q_{J} =
\frac{\lambda_{+} \, I_{0J}}{B_{+} (K-c_{+}^{(0)})^{2}} \, + \,
{\lambda_{+}\over\beta} \\
\upsilon \, Q_{c} =
\frac{\lambda_{+} \, I_{0c}}{B_{+} (K-c_{+}^{(0)})^{2}} \, + \,
\frac{\lambda_{+} \, (u_{M}-u_{1})}{\sqrt{(K-c_{+}^{(0)})^{2} +4\beta}}\\
\upsilon \, = \, \lambda_{+}^2 \, (u_{M}-u_{1}) \, - \,
B_{+} (K-c_{+}^{(0)})^{2}
\end{eqnarray*}

The velocity correction $c_{+}^{(1)}$ can be
obtained from the condition
$\partial P^{(1)}/\partial\chi =0$ at $\chi=\chi_m$. Thus
\begin{eqnarray}
Q + {1\over B_{+} (K-c_{+}^{(0)})^{2}}\int_0^{\chi_{m}}
e^{-(K-c_{+}^{(0)})\,\chi} P^{(0)}_{\chi} \nonumber\\
\times \left[P^{(0)}_{s} - J^{(1)}- c_{+}^{(1)}
P^{(0)}_{\chi} \right]\, d\chi = 0. \label{inner15}
\end{eqnarray}
Inserting (\ref{in3}) in this equation, we get $c_{+}^{(1)}$. After
simplification, the result is
\begin{eqnarray}
\gamma \, c_{+}^{(1)} = c_{+0} \, + \, c_{+J}
\, J^{(1)} _{c}
\label{inner16}
\end{eqnarray}
with
\begin{eqnarray*}
c_{+0} = \frac{J_{s}^{(0)}}{K-c_{+}^{(0)}} \left[
z_1 + \frac{z_2}{K-c_{+}^{(0)}}  
+ \left( \chi_m + z_2 \right)
\left( \frac{\lambda_+}{\beta} + \frac{1}{K-c_{+}^{(0)}} \right) \right] 
\\
+ \frac{c_{+\, s}^{(0)}}{(K-c_{+}^{(0)})^{2}}\left[
2 (\chi_m + z_2) \left(\frac{J^{(0)}-\alpha}{K-c_{+}^{(0)}} 
- \frac{\beta (u_{M}-u_{1})}{\sqrt{(K-c_{+}^{(0)})^{2} +4\beta}}
\right) \right. \\
\left. + (J^{(0)}-\alpha) \left( z_1 + \frac{z_2}{K-c_{+}^{(0)}}
\right) - \frac{B_+}{2} (K-c_{+}^{(0)})^{2} \chi_m^{2} \right]\\
- B_+ Q_{0} (K-c_{+}^{(0)})^{2} \\
c_{+J} = -z_2 - B_+ Q_{J} (K-c_{+}^{(0)})^{2} \\
\gamma = \frac{J^{(0)}-\alpha}{K-c_{+}^{(0)}} z_2 -
B_+ (K-c_{+}^{(0)}) \chi_m + B_+ Q_{c} (K-c_{+}^{(0)})^{2} \\
z_1 \equiv \frac{\chi_m \exp{[(K-c_{+}^{(0)}) \chi_m]}}
{(K-c_{+}^{(0)})} =
- \frac{\chi_m B_+ (K-c_{+}^{(0)})}{J^{(0)}-\alpha} \\
z_2 \equiv \frac{\exp{[(K-c_{+}^{(0)}) \chi_m]}-1}
{(K-c_{+}^{(0)})} = - \frac{ \lambda_+ (u_{M}-u_{1})}{J^{(0)}-\alpha}
\end{eqnarray*}
Here we have used (\ref{lambda}) - (\ref{chi_m}) to simplify the result. 
$J^{(1)}$ is found from the equation: 
\begin{eqnarray}
0 = 2 \int_{-\infty}^{0} (P^{(0)} - u_1)\, d\chi
+  {J^{(1)} - g'_{1} J_{s}^{(0)}\over g_{1}^{'\, 2}}\nonumber\\
+ 2 \epsilon \int_0^{\chi_{m}} P^{(1)}\, d\chi
+ \int_0^{\infty} (U_L(\xi)- u_1)\, d\xi \nonumber\\
+ \int_0^{\infty} (U_R(\xi)- u_1)\, d\xi,
\label{bias01}
\end{eqnarray}
after substitution of (\ref{bias0}), (\ref{bias1}) and (\ref{inner16}).

\renewcommand{\theequation}{C.\arabic{equation}}
\setcounter{equation}{0}
\section{Pulse Dynamics: Limiting cases}
\label{app-lim}
\subsection{Triangular pulse}
If the limiting pulse is triangular, the approximate evolution
equation for $J$ may be obtained by time-differentiating
(\ref{bias3}) and using (\ref{c+.limit}):
\begin{eqnarray}
{dJ\over dt} = - \sqrt{\epsilon}\, B(J)\, [K-
c_+(J)] \,, \label{J-1}\\
B(J) = \frac{4\beta (\beta u_{M}-\alpha)^{{1\over 4}} (\phi -
u_{1})^{{1\over 2}}\, (u_{M}- u_{1})^{{5\over 4}}}{2u_{M}  - \phi
- u_{1}}\,. \label{B}
\end{eqnarray}
For typical values of the parameters such as those  in Fig.\
\ref{fig1}, $B(J) >0$, so that $J$ tends to the solution of
$c_+(J)=K$. Triangular pulses are good approximations for $J$
sufficiently large, which means that the solution of (\ref{J-1})
decreases towards $J=\alpha$ according to (\ref{c+.limit}). Of
course before this value can be reached, (\ref{J-1}) ceases to be
valid and we revert to the general equation for $J$,
(\ref{eq_J1}), whose fixed point is $J^*$.

\subsection{Homoclinic pulse}
Let us now assume that $(K-c_+)\ll (J-\alpha)\ll 1$, whereas
$\chi_m \gg 1$. Then (\ref{B+}) and (\ref{chi_m}) imply that
\begin{eqnarray}
\chi_m \sim {\lambda_{+}\, (u_{M} - u_{1})\over J -\alpha} +
{\lambda_{+}^{2} (u_{M} - u_{1})^{2}(K-c_{+})\over 2 (J
-\alpha)^{2}}\,. \label{chi_m1}
\end{eqnarray}
Here $\lambda_+ \sim \sqrt{\beta}$. Notice that (\ref{chi_m1})
becomes (\ref{zeta_0.homo}) as $(K-c_+)\to 0$. We now insert this
approximation in (\ref{u_m}) after (\ref{B+}) has been
substituted. The result is
\begin{eqnarray}
u_m \sim u_M + {\lambda_{+}^{2} (u_{M} - u_{1})^{2}\over 2 (J
-\alpha)} \sim {1\over 2}(J -\alpha) \chi_m^2 . \label{u_m1}
\end{eqnarray}
The bias condition (\ref{bias1}) may now be approximated by using
(\ref{chi_m1}) and (\ref{u_m1}) to obtain
\begin{eqnarray}
{\phi -u_{1}\over \epsilon} \sim {2 \beta^{{3\over 2}}
(u_{M}-u_{1})^{3}\over 3 (J-\alpha)^{2}}\,.\nonumber
\end{eqnarray}
Then
\begin{eqnarray}
(J-\alpha) &\sim & \beta^{{3\over 4}} \sqrt{ {2\epsilon
(u_{M} - u_{1})^{3}\over 3 (\phi - u_{1})} }\label{estimates2}\\
&\sim & \beta^{{3\over 4}} \sqrt{ {2\epsilon \left( u_{M} -
{\alpha\over\beta}\right)^{3}\over 3 \left(\phi - {\alpha
\over\beta}\right)} }\,. \nonumber
\end{eqnarray}
Inserting (\ref{estimates2}) in (\ref{chi_m1}) and (\ref{u_m1}),
we obtain
\begin{eqnarray}
\chi_m \sim \beta^{- {1\over 4}} \sqrt{ {3 (\phi - u_{1}) \over
2\epsilon\, (u_{M} - u_{1})} }\,,
\label{estimates1}\\
u_m \sim {\beta^{{1\over 4}}\over 2}\,\sqrt{{3\over 2\epsilon}\,
(\phi - u_{1})\, (u_{M} - u_{1}) }\,.  \label{u_m2}
\end{eqnarray}
Equations (\ref{estimates2}) - (\ref{u_m2}) are the same as
(\ref{J.homoclinic}) - (\ref{u_+.homoclinic}) for the homoclinic
pulse.

As explained before, $d\chi_m/dt = (c_- -c_+)/2 = K-c_+$.
Therefore the derivative of (\ref{estimates1}) with respect to
time yield
\begin{eqnarray}
{dJ\over dt} \sim - \sqrt{{8\epsilon\over 3}}\, {\beta^{{5\over
4}}\, (\phi - u_{1})^{{1\over 2}} (u_{M} - u_{1})^{{3\over
2}}\over u_{M} - \phi }\, (K-c_{+})\,.   \label{eq_J2}
\end{eqnarray}
This equation has the same form as (\ref{eq_J1}), and it shows
that the unknown $J(t)$ varies on a slow time scale $t =
O(1/\sqrt{\epsilon (K-c_{+})}$. In the present limit, $(K-c_{+})
\ll (J-\alpha) = O(\sqrt{\epsilon})$, so that the corresponding
time scale is slower than $\tau=\epsilon t$.
\bigskip

A glance to Eq.~(\ref{eq_J2}) shows that $J$ decreases
exponentially fast to $J^*$ such that $c_+ = c_- = K$. The
resulting pulse is the homoclinic orbit of the phase plane
(\ref{fronts}) with $c=K$ described in the
previous Section.

\end{document}